\documentclass[reprint,amsmath,amssymb,aps,pre]{revtex4-1}

\usepackage{graphicx}
\usepackage{dcolumn}
\usepackage{bm}

\usepackage[breaklinks=true]{hyperref}

\hypersetup{colorlinks=true, linktoc=all, linkcolor=blue, linktocpage, citecolor=blue}

\newcommand{\overbar}[1]{\mkern 1.5mu\overline{\mkern-1.5mu#1\mkern-1.5mu}\mkern 1.5mu}

\begin{document}

\title{Phase Transitions in Hardcore Lattice Gases on the Honeycomb Lattice}

\author{Filipe C. Thewes}
\email{filipe.thewes@ufrgs.br}
\author{Heitor C. M. Fernandes}%
 \email{heitor.fernandes@ufrgs.br}
\affiliation{%
 Instituto de F\'\i sica, Universidade Federal do Rio Grande do Sul - CP 15051  91501-970, Porto Alegre, RS, Brazil\\
}%

\date{\today}

\begin{abstract}
  We study lattice gas systems on the honeycomb lattice where particles exclude neighboring sites up to order $k$ ($k=1\ldots5$) from being occupied by another particle.  Monte Carlo simulations were used to obtain phase diagrams and characterize phase transitions as the system orders at high packing fractions. For systems with first neighbors exclusion (1NN), we confirm previous results suggesting a continuous transition in the 2D-Ising universality class.  Exclusion up to second neighbors (2NN) lead the system to a two-step melting process where, first, a high density columnar phase undergoes a first order phase transition with non-standard scaling to a solid-like phase with short range ordered domains and, then, to fluid-like configurations with no sign of a second phase transition. 3NN exclusion, surprisingly, shows no phase transition to an ordered phase as density is increased, staying disordered even to packing fractions up to 0.98. The 4NN model undergoes a continuous phase transition with critical exponents close to the 3-state Potts model. The 5NN system undergoes two first order phase transitions, both with non-standard scaling. We, also, propose a conjecture concerning the possibility of more than one phase transition for systems with exclusion regions further than 5NN based on geometrical aspects of symmetries.
\end{abstract}

\maketitle


\section{\label{sec:intro}Introduction}

Lattice systems are one of the main tools in studying phase transitions and critical phenomena in statistical physics. Composed of particles occupying lattice sites and interacting with their vicinity as well as external fields, these systems are of great importance in understanding the influence of symmetries in phase transitions~\cite{runnelsPT}. 
First introduced as a discrete version for the problem of hard spheres~\cite{alderHardSpheres,dickmanHS}, a well established problem is the hardcore lattice gas~\cite{Burley1960}, where the only interaction considered is the prohibition of a given region around a particle from being occupied by another particle. 
In this case, with a suitable choice of the underlying lattice and the excluded region, it is possible, at least in principle, to develop a hardcore lattice model for almost any particle shape, which determines the full packing configurations and all different phases occurring as density is decreased~\cite{runnelsPT,frenkelEntropy}. Moreover, since hardcore interactions are athermal, every phase transition is entropy driven, with ordered phases showing higher entropy than disordered ones \cite{frenkelEntropy}.

Given their simplicity and wide coverage of underlying symmetries, hardcore lattice gases allow us to study several different classes of phase transitions and critical behavior, including freezing transitions~\cite{Pusey1986}, polymer induced attraction in colloidal particles~\cite{asakura1958} and phase separation in binary mixtures~\cite{frenkel1992}. For this reason, studying these models, as well as any other toy model, is a means of exploring the field of statistical mechanics in search of interesting phenomena, leading to insights about novel forms of experiments, technologies, and theories. 

While some systems have approximate results obtained by means of analytical procedures~\cite{runnels1HC,verberkmoesTriang,lafuente,juergen2011},
only the hard hexagons model has an exact solution~\cite{baxter_hh}. Monte Carlo simulations have been used to study several other particle shapes and their mixtures,~\cite{heitor,ramola2012,rajesh,yShapedRaj,dimers,rodsRaj,heitor2,rajeshcubes,panagiotopoulos,dickman2012}, most of them on the square, triangular or cubic lattices. Recently, a number of models with both symmetrical and asymmetrical particles have been studied in continuous~\cite{pagonabarraga} and discrete~\cite{rajesh,ramola2015,KunduRectangles} space, showing several phase transitions, including high density columnar phases missed by earlier studies on the square lattice and an hexatic phase on the triangular lattice~\cite{darjani_2019}. 
In contrast to previous approaches, which employed single particle modifications during sampling, recent studies employ highly efficient cluster algorithms~\cite{yShapedRaj,rajesh,rajeshcubes,rodsRaj,ramola2015,KunduRectangles}, enabling simulations of systems with up to $N=1024^2$ lattice sites. Another very interesting result is the possibility of multiple phase transitions for larger exclusion regions whenever a sliding instability is present at high density phases~\cite{rajesh,yShapedRaj,Nath_2016}.
Finally, the melting of 2D materials has also attracted attention lately, undergoing many interesting critical phenomena~\cite{2dMelting, hardPolygons, 2dSpheres}. 

In this paper, we investigate the hardcore model on the honeycomb lattice, where neighboring sites up to order $k$ ($k$NN) of a particle are prohibited from being occupied by another particle. While some models with finite interaction on nearest/next nearest neighbors on the triangular~\cite{santi2000,santi2007} and honeycomb~\cite{devilStep3nn,kanamoriHC,diffusionHC,isingAFHC} lattices have been studied, an extensive and systematic investigation of the hardcore interaction on the honeycomb lattice is still lacking.
Here, we perform simulations for $k$ up to $5$ and, using finite size scaling methods, we characterize the different phase transitions these models undergo. We also develop a conjecture for some $k$ where we expect at least one sublattice phase transition
and check this for $k=9,11\text{ and }14$. Our conjecture also hints at possible phases for higher $k$, including hexatic and columnar phases. 

The most challenging aspect of the honeycomb lattice is its ability to accommodate very stable high density domain like phases, in which there is local but not global order.
Transitions between such phases and ground state configurations require highly efficient sampling algorithms and may become rare for larger systems.

The remainder of this paper is structured as follows. In Sec.~\ref{sec:model} we introduce the model and briefly describe the grand canonical Monte Carlo algorithm used. We present results and discuss the different phases found in Sec.~\ref{sec:results}. In Sec.~\ref{sec:conjec} we develop a conjecture for higher values of $k$ and summarize our results in Sec.~\ref{sec:summary}.

\section{\label{sec:model}Model and Algorithm}

\begin{figure}[hbt] 
  \begin{center}
    \includegraphics[width=0.48\columnwidth]{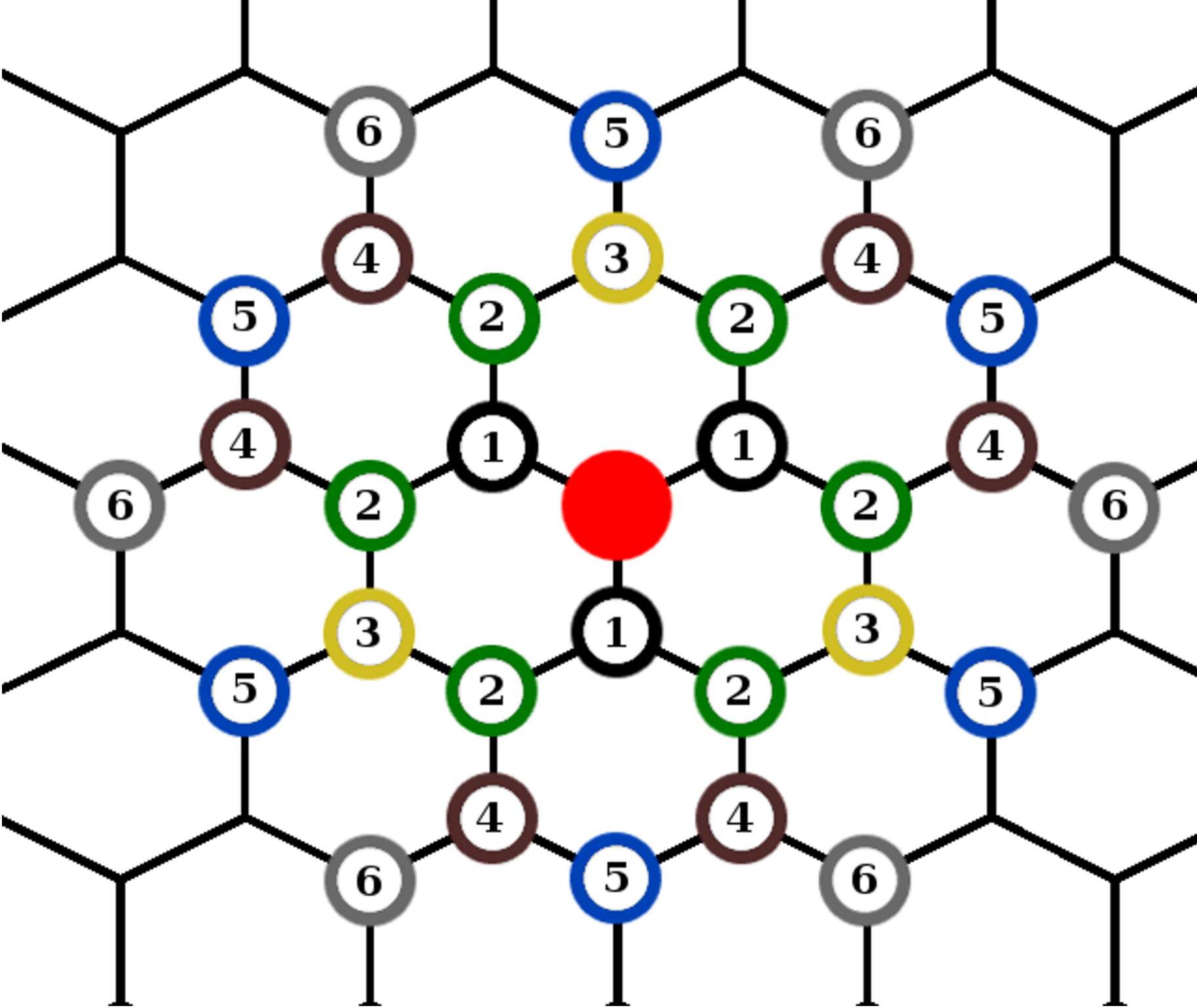}
    \includegraphics[width=0.48\columnwidth]{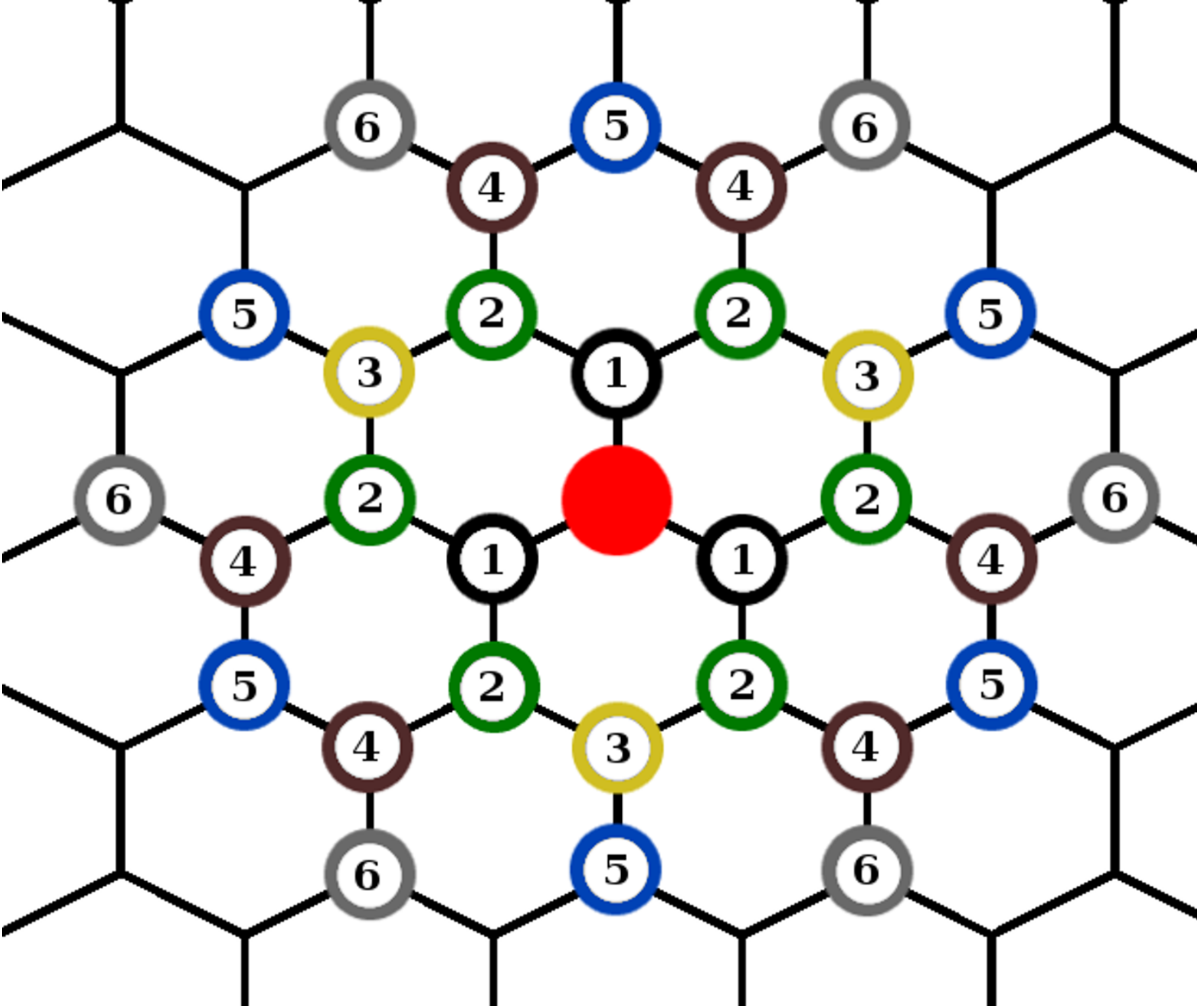}
    \end{center}
    \caption{For a given orientation of the whole lattice, two types of sites (left and right) are present on the honeycomb lattice. A particle (center) has three first, six second, three third, six fourth, six fifth and six sixth nearest neighbors.}
    \label{exclusion}
\end{figure}

A $k$NN hardcore gas model on a lattice is a system in which particles occupy lattice sites (vertices) and prohibit its neighbors 
of order up to $k$ of being occupied by another particle. In the grand canonical ensemble, an activity $z=e^\mu$ is assigned to particles, where $\mu$ is the chemical potential. Figure~\ref{exclusion} shows the exclusion up to $k=6$ for the honeycomb lattice.

\begin{figure}[thb] 
  \begin{center}
    \includegraphics[width=.8\columnwidth]{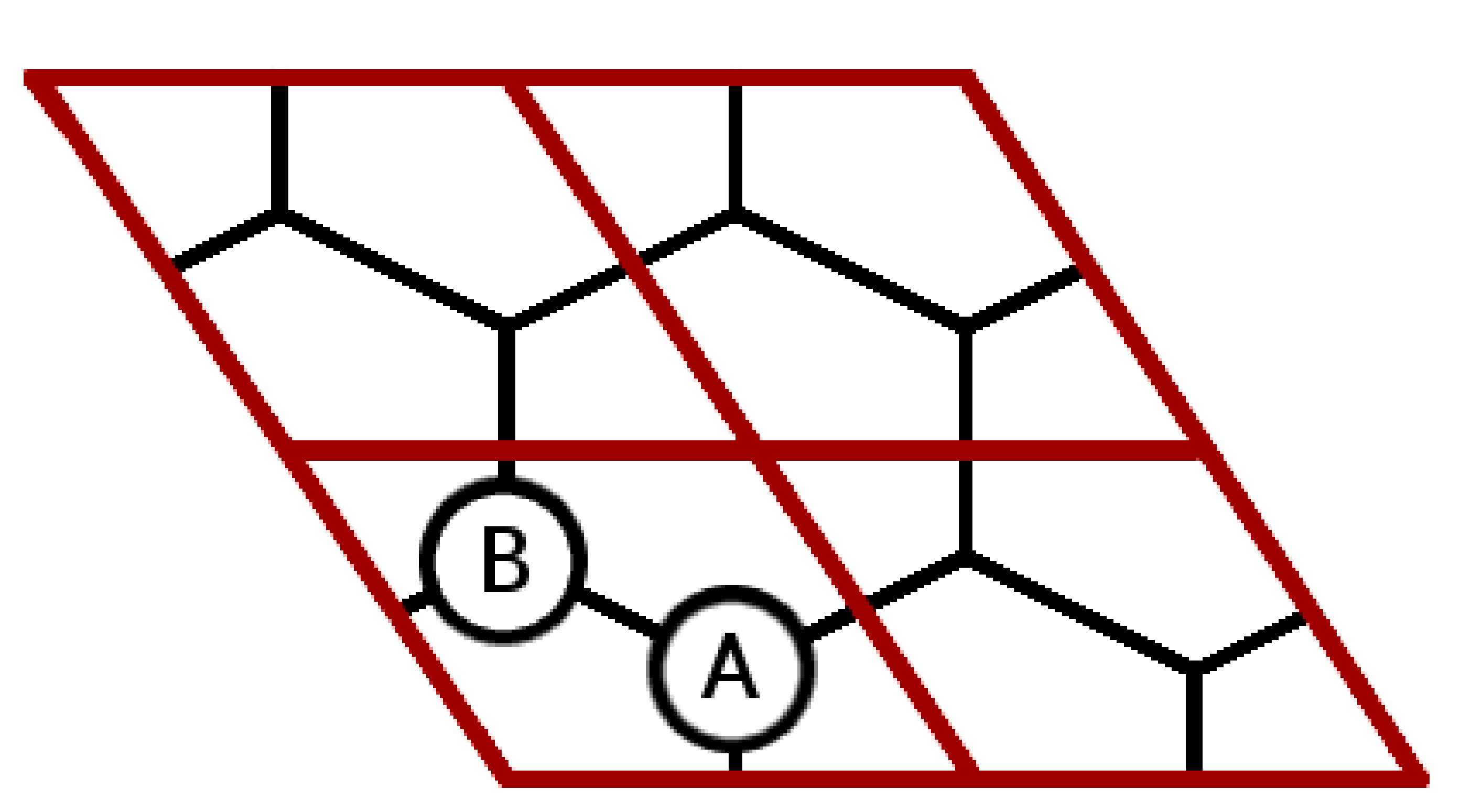}
    \caption{The lattice is defined as an $L\times L$ square grid of unitary cells, each containing one site of type $A$ and one of type $B$. Total number of sites is $2L^2$.} 
    \label{lattice}
  \end{center}
\end{figure}

Since the honeycomb lattice is composed of two superimposed triangular lattices, which we call lattices $A$ and $B$, we define it as an $L\times L$ tilted square grid of unitary cells, each one containing one $A$-site and one $B$-site. Periodic boundary conditions are imposed along the two directions of the square grid. Figure~\ref{lattice} shows this lattice construction.

It is well known that when studying systems at high density, or area fraction, algorithms using single particle movements such as the grand canonical Metropolis or multicanonical Wang-Landau are inefficient at densities close to full packing~\cite{jankeMUCA}. Therefore, we use an efficient cluster algorithm introduced in Ref.~\cite{rajesh} to study the $k$NN model on the square lattice which has shown to be able to equilibrate volume (area, in 2D) fractions up to $0.99$ in a system of hard rods on the square lattice~\cite{rodsRaj}.

We briefly describe the algorithm adapted for the honeycomb lattice. First, one of the two triangular lattices (say $A$) is randomly selected. In this lattice, a row is chosen and one of the three possible lattice directions is picked. All particles along this $A$-row are evaporated (deleted) and the row now consists of intervals of sites able to be populated separated by blocked sites due to particles on neighboring rows as well as particles on the $B$ lattice. The reoccupation of these intervals is reduced to a 1D $q$-mer problem, with well known equilibrium probabilities, ensuring that balance condition is satisfied. Ergodicity condition is more subtle and we refer to Ref.~\cite{ramola2015} for a more detailed discussion. 
A Monte Carlo movement is completed after updating $6L$ rows. Since this algorithm is easily parallelizable, we use an OpenMP~\cite{openMP} version where multiple rows distant of at least $\Delta$ (Table~\ref{slidingS}) are simultaneously updated in the same direction. We use the PCG~\cite{pcgPRNG} pseudo-random number generator. 

\begin{figure}
    \begin{center}
        \includegraphics[width=.7\columnwidth]{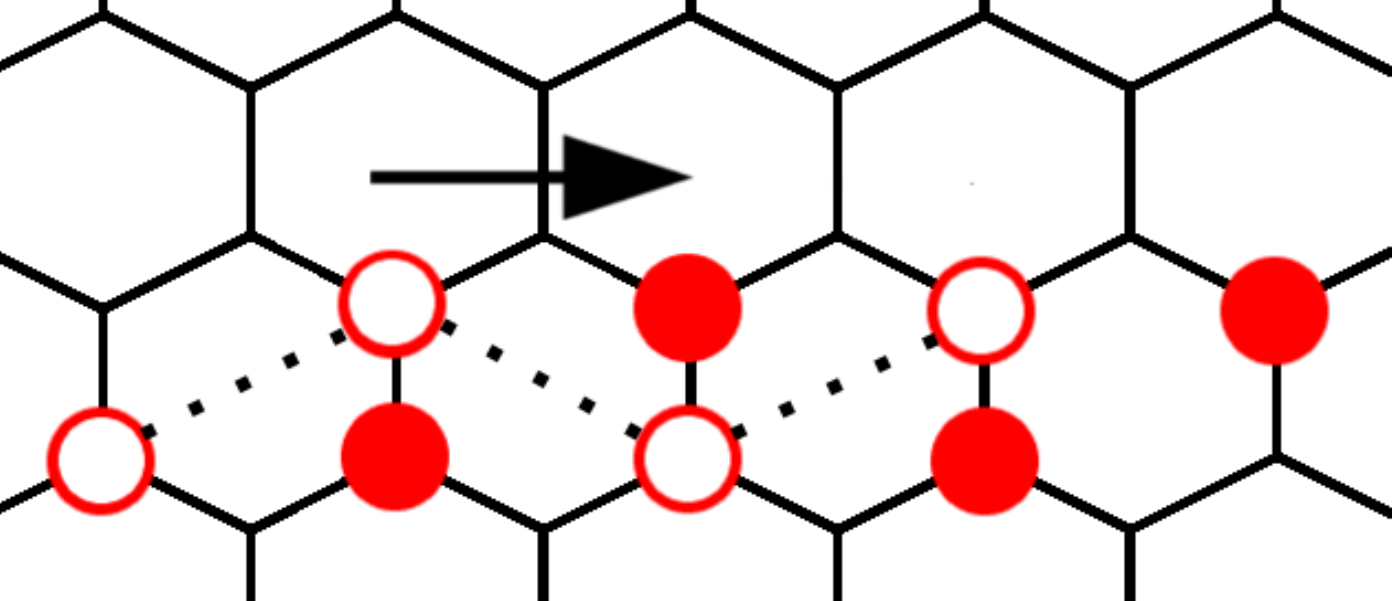}
        \caption{Cluster formation (dotted lines) and sliding movement for the 2NN case. Empty circles form a cluster and are moved in the arrow direction. Filled circles show the final position of particles in the cluster. By symmetry, there are six possible directions for cluster formation.}
        \label{slidingClust}
    \end{center}
\end{figure}

Our results show that, even with cluster movements, systems under study do not explore phase space very efficiently. Therefore, in order to improve our sampling, in the 2NN case we proceed as in~\cite{yShapedRaj} and add a sliding movement in which a linear cluster is formed and slid in a given direction (see Fig.~\ref{slidingClust} for illustration). To form a cluster, a root particle and one of the six directions are randomly picked. As long as the next site in the given cluster direction is occupied, particles are added to this cluster. It should be noted that particles lying on both $A$ and $B$ lattices are used to build the cluster. A sliding movement is performed if it does not violate the hardcore constraints. For this kind of trial movement, detailed balance is clearly obeyed since the reverse movement, that is, choosing the last particle as root and building the cluster in the opposite direction occurs with same probability. In cases 4NN and 5NN we perform single particle movements instead of cluster sliding. To achieve this, a particle and a site able to be occupied are randomly selected and the particle is moved into that site. In order to perform this movement efficiently, we keep track of both particles and free sites during the simulation. In all cases, the canonical movement is performed $2L^2/S$ times every Monte Carlo step (see Table~\ref{slidingS} for the values of $S$ used in each case). We choose an efficient $S$ but did not investigate optimal choices.

\begin{table}[h]
  \begin{center}
    \begin{tabular}{  c || c | c | c | c | c | }
        kNN  & \hspace{.3cm}1\hspace{.3cm} & \hspace{.3cm}2\hspace{.3cm} & \hspace{.3cm}3\hspace{.3cm} & 
      \hspace{.3cm}4\hspace{.3cm} & \hspace{.3cm}5\hspace{.3cm} \\ \hline
      $\Delta$ & 3 & 3 & 3 & 5 & 5\\
      $S$ & - & 5 & - & 10 & 5\\
      \hline
    \end{tabular}    
    \caption{Values of $\Delta$ and $S$ for the different cases. We update simultaneously rows distant of $\Delta$ from each other and perform $2L^2/S$ canonical movements every Monte Carlo step.}
    \label{slidingS}
  \end{center}
\end{table}

\section{\label{sec:results}Results}

\subsection{\label{subsec:1nn}Nearest neighbors exclusion ($k=1$).}

The case where $k=1$ undergoes a phase transition from a low density, fluid-like phase, to a high density, solid-like phase, as chemical potential is increased. This transition  is expected to belong to the 2D Ising universality class, as pointed in Refs.~\cite{runnels1HC,debierre1HC}.

\begin{figure}[thb] 
  \begin{center}
    \includegraphics[width=8cm]{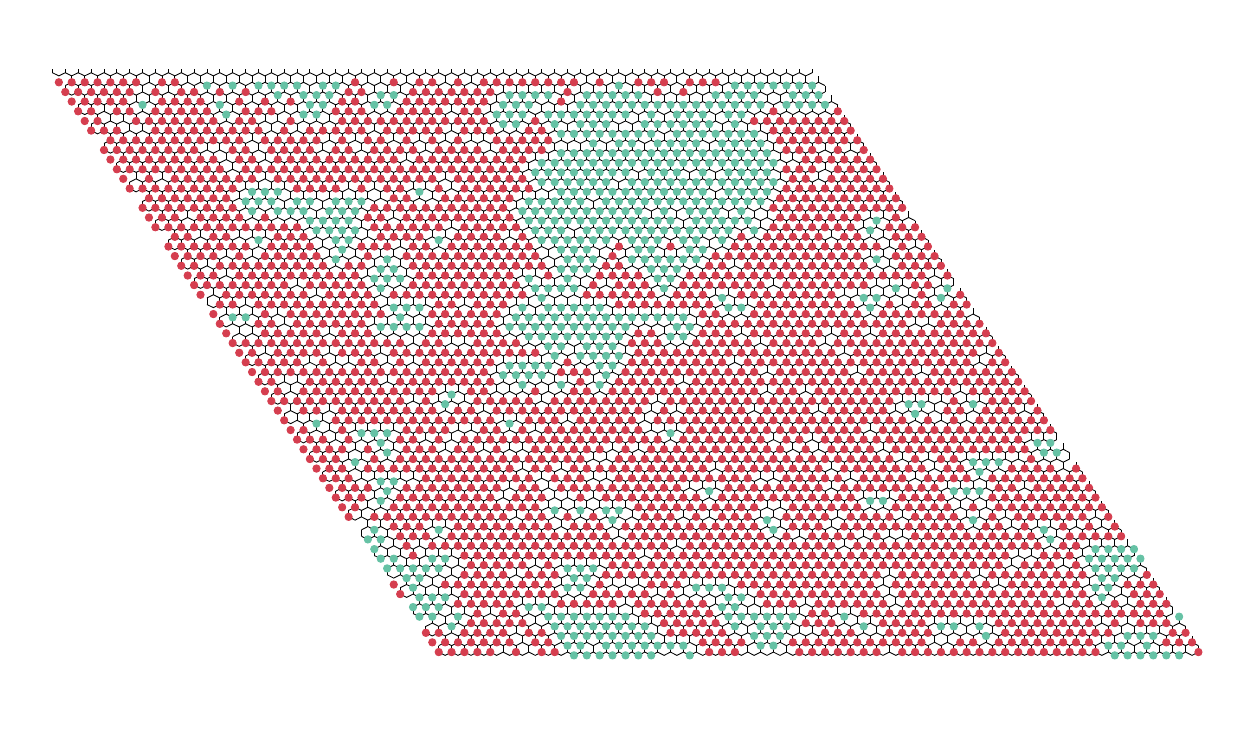}
    \caption{Snapshot of a typical configuration near the phase transition ($\mu=2.05$) for the 1NN case. $L=60$ and $\rho\simeq 0.415$. It is possible to see both types of sites, in different colors, present but no sign of coexistence of a disordered and an ordered phase, indicating a second order transition.} 
    \label{snap1nn}
  \end{center}
\end{figure}

For lower densities, we observe a disordered phase with symmetric occupation of $A$ and $B$ sites. As density is increased, a spontaneous symmetry breaking takes place at critical chemical potential $\mu_c=2.064$ (see snapshot in Fig.~\ref{snap1nn}) and one sublattice ($A$ or $B$) is preferentially occupied. We study this phase transition using the algorithm described in Section~\ref{sec:model}. In this case, evaporation and deposit of particles is already highly efficient and we do not perform canonical movements.

To characterize the phase transition, we define an order parameter as
\begin{equation}
  Q_1 = 2|\rho_A - \rho_B|,
\end{equation}
where $\rho_i$ denotes the density of sites of type $i$ and factor 2 takes into account that the maximum density possible is $1/2$.  

We also measure the susceptibility $\chi_1$ of the order parameter
\begin{equation}
  \chi_1 = 2L^2(\langle Q_1^2 \rangle - \langle Q_1 \rangle^2) \,.
\end{equation}

Whenever there is no risk of confusion, in the remainder of this paper we omit the ensemble (time) average symbol $\langle Q \rangle$ in favor of only $Q$.

\begin{figure}[thb] 
  \begin{center}
    \includegraphics[width=0.8\columnwidth]{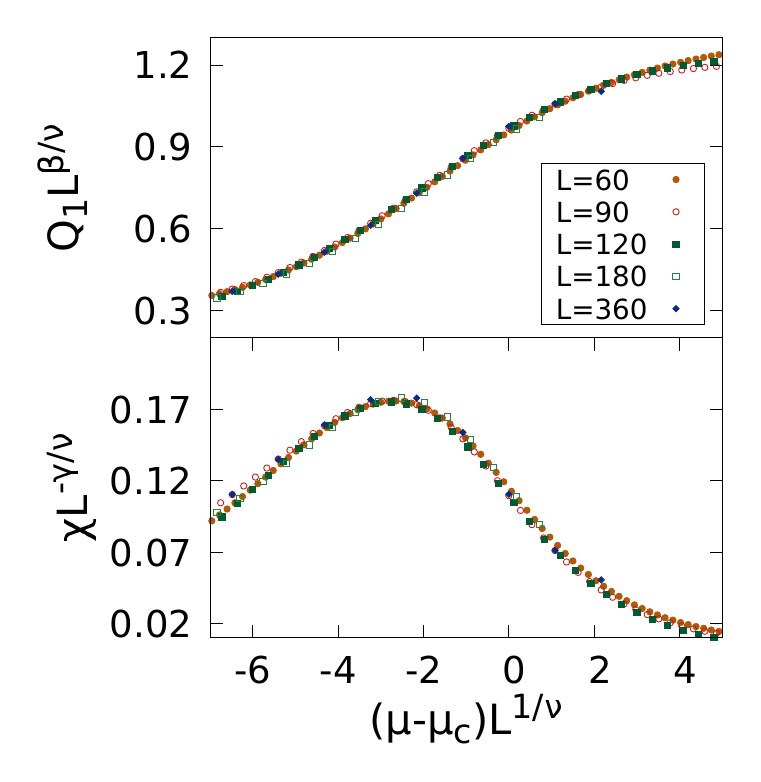}
    \caption{Finite size scaling collapse of curves of order parameter $Q_1$ (top) and its susceptibility (bottom) for the 1NN case for different $L$. We use the Ising-2D critical exponents $\gamma=7/4$, $\beta=1/8.$ and $\nu=1$, as suggested in previous results~\cite{debierre1HC,runnels1HC}.}
    \label{col1nn}
  \end{center}
\end{figure}

After performing a long simulation near the critical point, we use the histogram re-weighting technique to extrapolate data~\cite{hrw88}. Figure~\ref{col1nn} shows the collapsed curves for these quantities for different system's size $L$ after re-scaling using the finite size theory. 

We confirm the results previously obtained in~\cite{debierre1HC,runnels1HC}, with a transition at $\mu=2.064$ and critical exponents in the Ising-2D universality class.

\subsection{Up to second neighbors exclusion ($k=2$)}
\label{subsec:2nn}

\begin{figure}[thb] 
  \begin{center}
    \includegraphics[width=6cm]{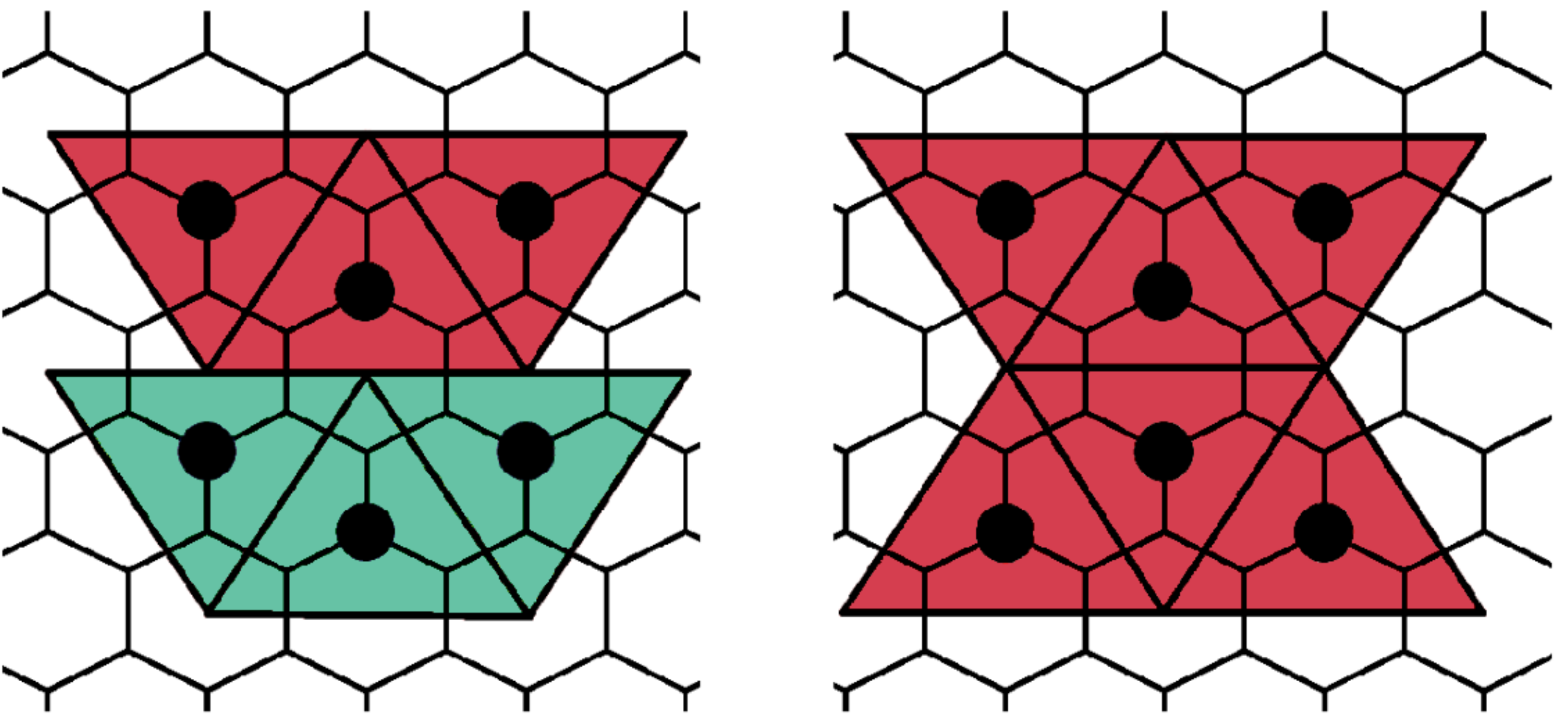}
    \caption{Primitive cell of particles and two possible full packing configurations for the 2NN model. Since particle shape can be seen as
    an equilateral triangle, a translation of one unit cell along the three lattice directions is possible without compromising the full packing.}
    \label{ws2nn}
  \end{center}
\end{figure}

We start by constructing one of the possible primitive cells for the system, where particles have an equilateral triangle shape (Fig.~\ref{ws2nn}). In our model, a row of closed packed triangles of size $2\times L$ can be slid by one lattice unit without compromising the full packed configuration and a second sliding brings the row to its initial state. Thus, each row has two possible states at maximum density. Since the honeycomb lattice has three equivalent directions where such rows can be formed, the ground state of this model has a $6\times 2^{L/2}-3$ degeneracy.

\begin{figure}[thb] 
  \begin{center}
    \includegraphics[width=0.7\columnwidth]{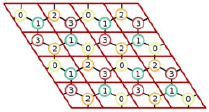}
    \caption{Sublattice definitions for the 2NN model.
    }
    \label{sl2nn}
  \end{center}
\end{figure}

To account for the sliding freedom, we define four sublattices as depicted in Figure~\ref{sl2nn} and calculate the following quantities:

\begin{equation}
  \label{pos2nn}
  \begin{split}
    q_0 & = 4| \rho_0 + \rho_2 - \rho_1 - \rho_3 | \\
    q_+ & = 4| \rho_0 + \rho_1 - \rho_2 - \rho_3 | \\
    q_- & = 4| \rho_0 + \rho_3 - \rho_1 - \rho_2 |.
  \end{split}
\end{equation}

Each of the components in equation~(\ref{pos2nn}) measures ordering along one of the lattice directions. The factor $4$ takes into account that the maximum density, at close packing, is $1/4$. 

To quantify the phase transition we define the order parameter $Q_2$ as
\begin{equation}
  Q_2 = \textrm{max}(q_0,q_+,q_-),
  \label{eq:q2nn}
\end{equation}
where function $\max(x,y,z)$ returns the greatest value of its arguments.

\begin{figure}[thb] 
  \begin{center}
    \includegraphics[width=0.9\columnwidth]{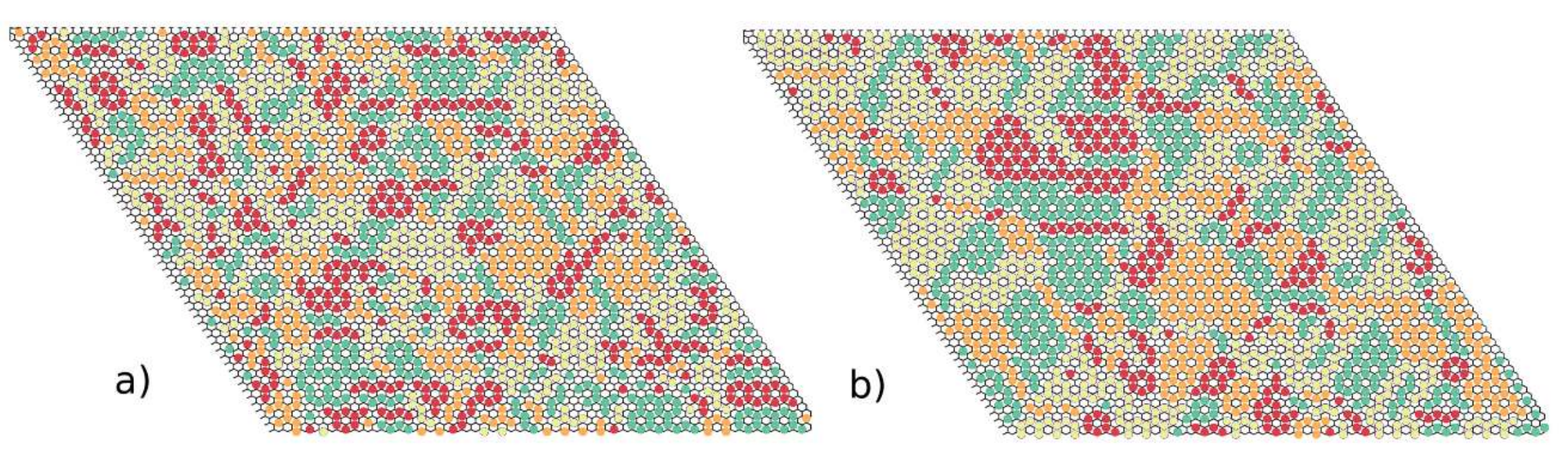}
    \includegraphics[width=0.9\columnwidth]{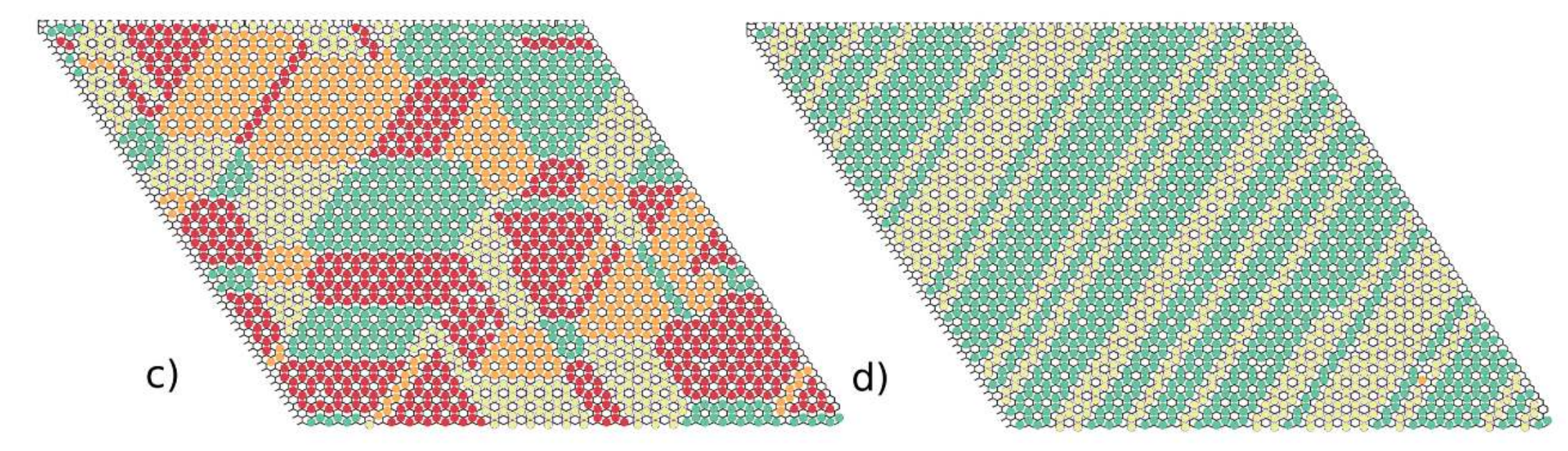}
    \caption{Typical configurations for the 2NN case. Snapshots are for $L=60$ at $\mu$[$\rho$]: (a) $3.0$[$0.211$], (b) $4.0$[$0.226$], (c) $5.2$[$0.244$] and (d) $5.5$[$0.248$]. Colors (shades of gray) show sublattices as defined in Fig.~\ref{sl2nn}. In panel (a) there is no ordering and the system is in a fluid-like configuration. As chemical potential is increased, panels (b) and (c) show ordered domains that remain stable in size at fixed $\mu$ but grow as $\mu$ is increased. Finally, panel (d) shows how the system breaks into independent slabs after the phase transition related to order parameter $Q_2$ occurs and a symmetry break in the occupancy of sublattices is observed.}
    \label{snap2nn}
  \end{center}
\end{figure}

In Fig.~\ref{snap2nn} we show snapshots of typical configurations for different $\mu$. We find a phase transition from a domain-like phase into a full packing configuration where the system breaks into independent slabs of size $2\times L$.

\begin{figure}[thb] 
  \begin{center}
    \includegraphics[width=0.98\columnwidth]{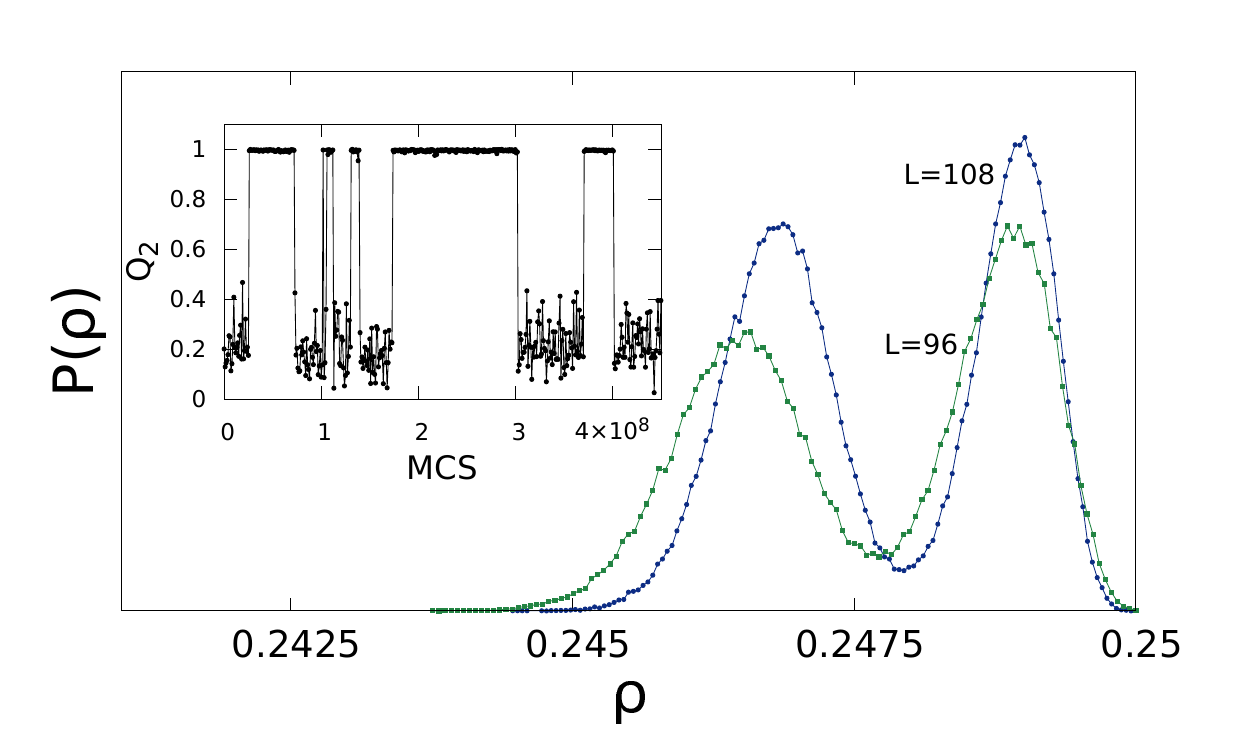}
    \caption{Histograms of density for $L=96$ and $L=108$ in the 2NN case show two peaks, indicating a first order phase transition. Inset: part of time series of order parameter $Q_2$ for $L=108$ at $\mu=5.83$. We stress the scale on the time-axis ($10^8$ MCS). Due to the long time it takes to
    jump between phases, sampling even small system sizes as $L=108$ becomes very difficult.}
    \label{hist2nn}
  \end{center}
\end{figure}

Since probability distributions (histograms) of both density and order parameter $Q_2$ show two peaks (Fig.~\ref{hist2nn} and inset, respectively), we expect a first order phase transition to occur. 

\begin{figure*}[!htb] 
  \begin{center}
    \includegraphics[width=0.32\textwidth]{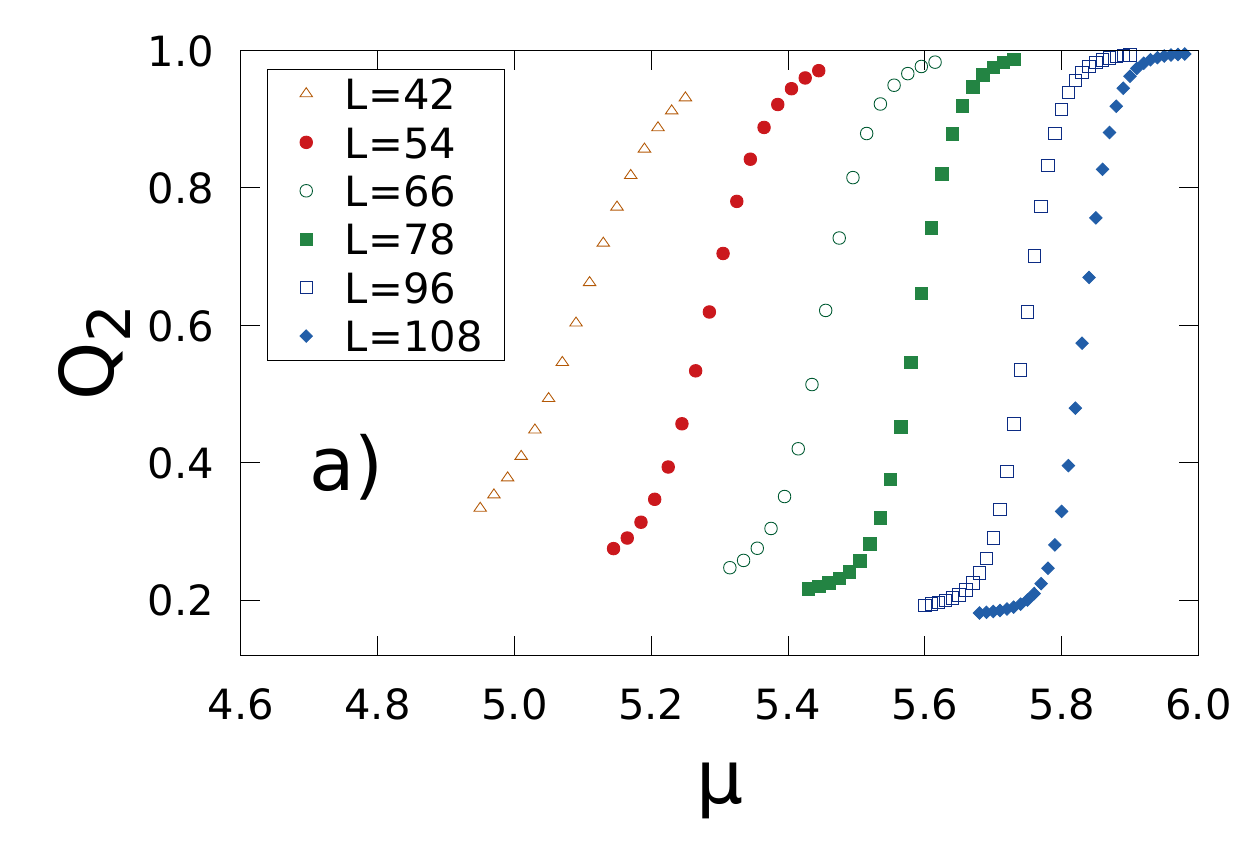}
    \includegraphics[width=0.32\textwidth]{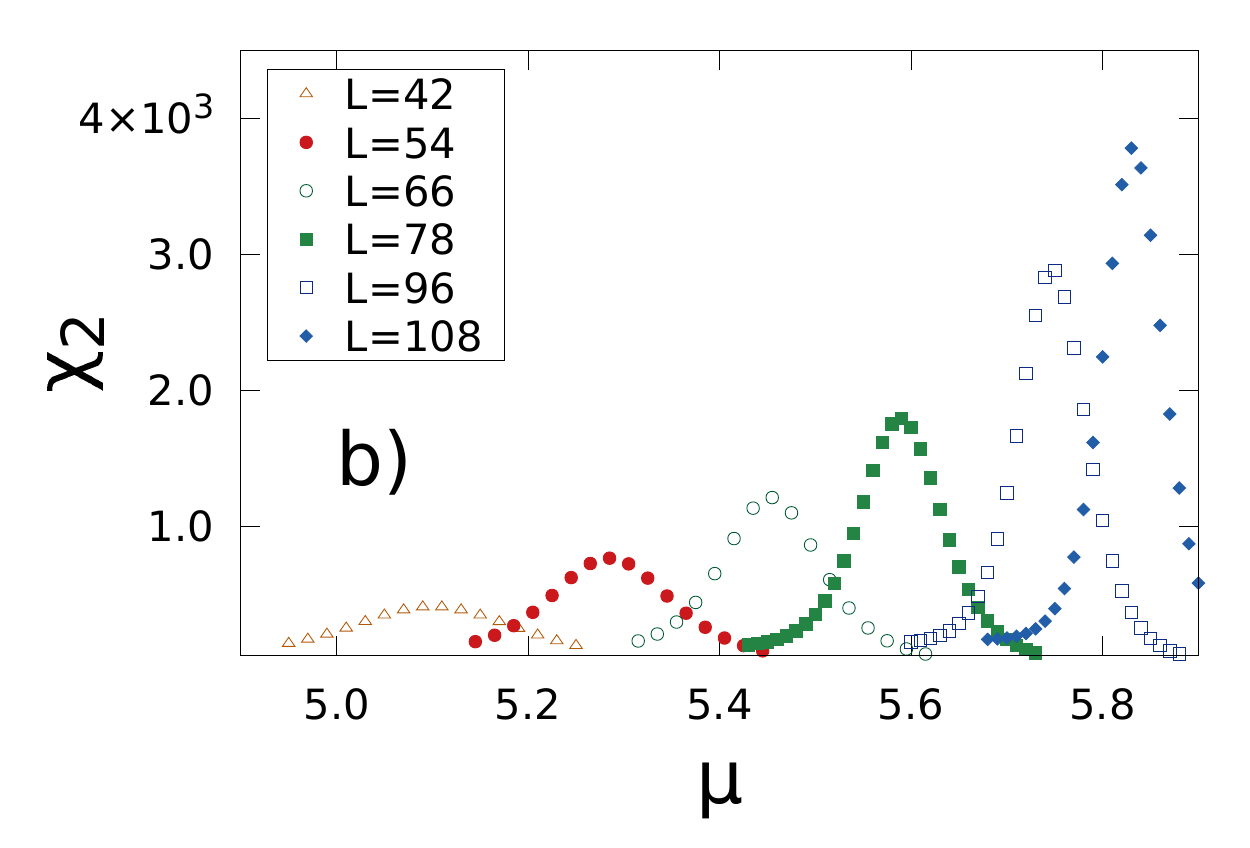}
    \includegraphics[width=0.32\textwidth]{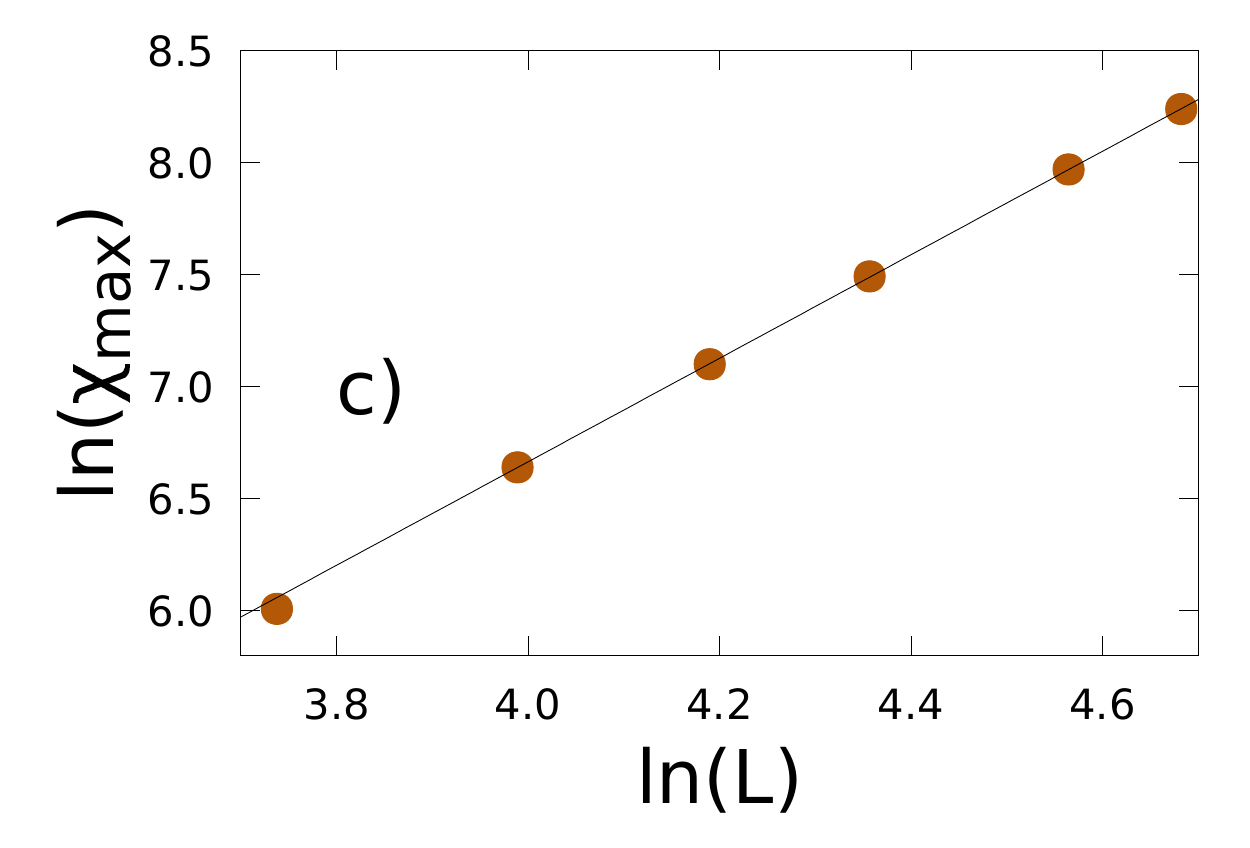}
    \includegraphics[width=0.32\textwidth]{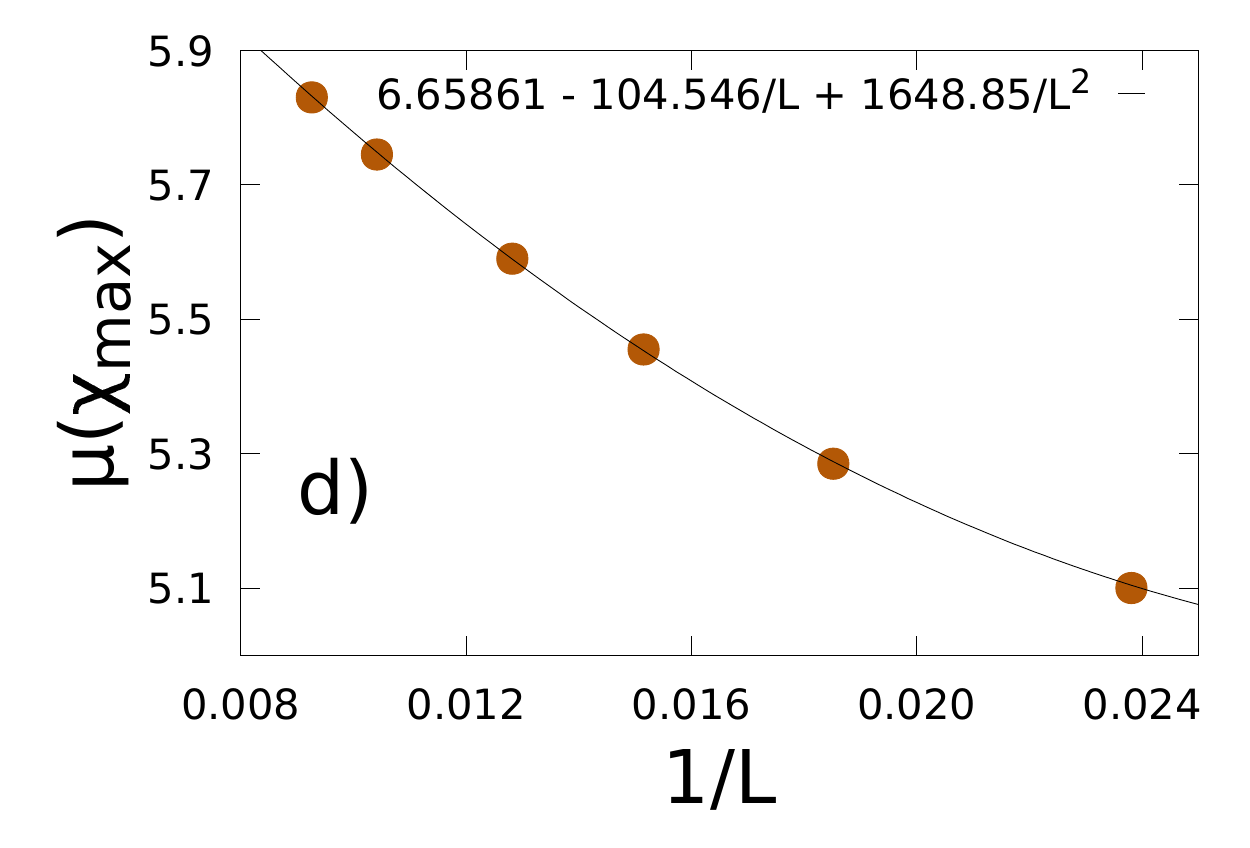}
    \includegraphics[width=0.32\textwidth]{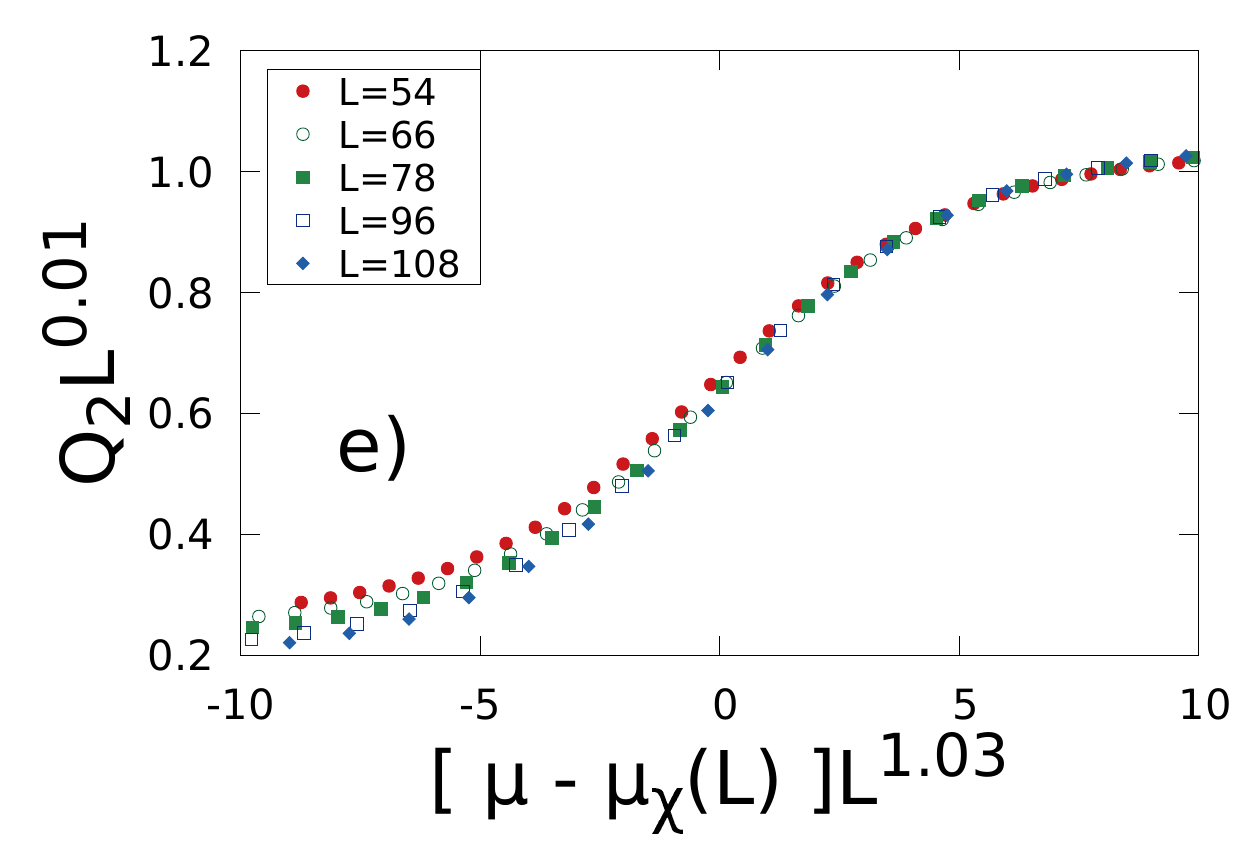}
    \includegraphics[width=0.32\textwidth]{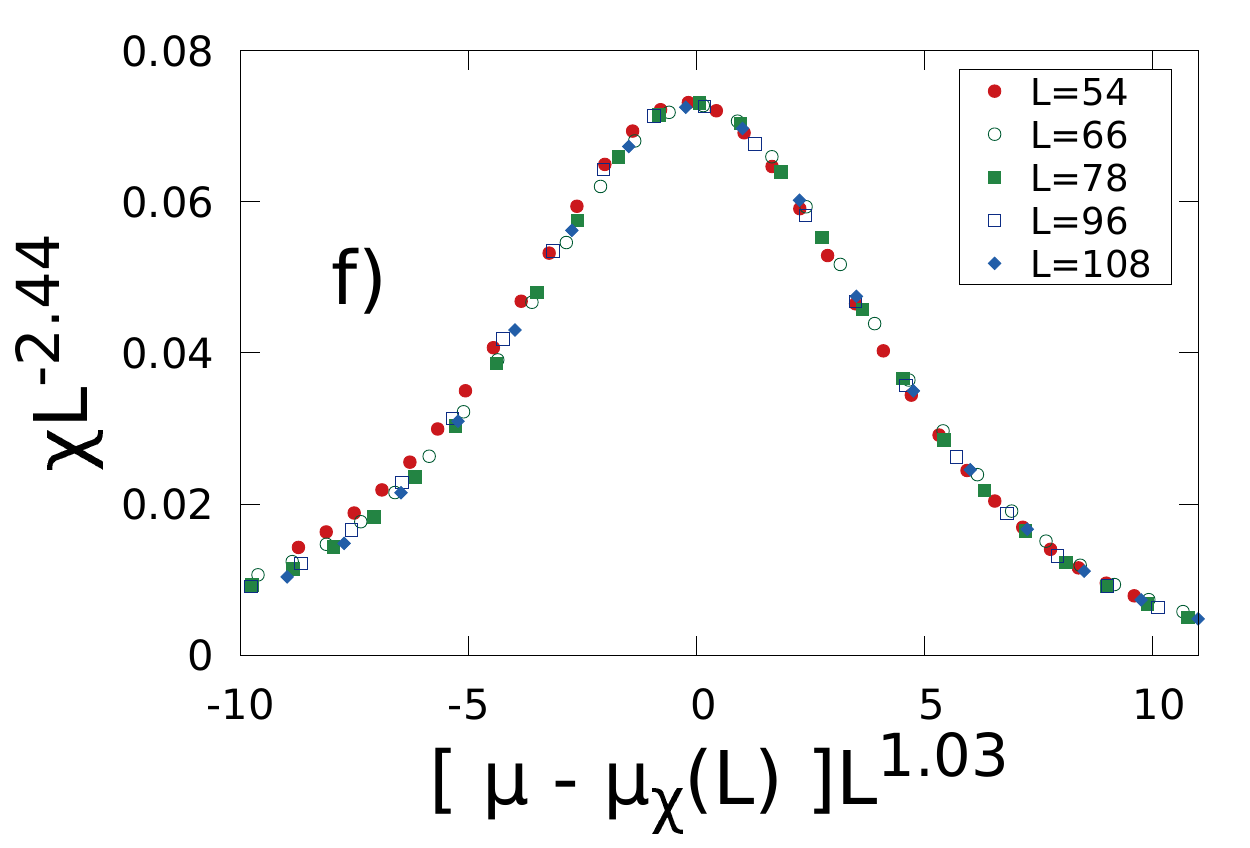}
    \caption{Results for the 2NN case. From top-left to bottom-right: (a) Order parameter $Q_2$ and (b) its susceptibility $\chi_2$ as function of chemical potential, $\mu$, for different sizes, $L$. (c) Dependence of $\chi_{max}$ with $L$. The solid line has slope $2.44$. In bottom line, (d) non-standard scaling as discussed in relation (\ref{scalingF}). For large $L$, the position of the first order phase transition scales with $L$ instead of the standard scaling with $L^2$. See text and references for discussion. Finite size scaling collapse of (e) order parameter and (f) susceptibility for different lattice sizes $L$ with non-standard scaling exponents.}
    \label{qmu2nn}
  \end{center}
\end{figure*}

In Ref.~\cite{JankeNonstandard} authors argue that, in first order phase transitions where ground state degeneracy grows exponentially with system size ($\sim 2^{L/2}$, in our case), standard scaling laws must be modified and quantities such as $\mu_c(L)$ do not scale with $L^d$ but with $L^{d-1}$, where $d$ is system dimension.

Therefore, we proceed as in Ref.~\cite{IsingPlaquettes} and adjust values of critical chemical potential obtained by the maximum of the order parameter susceptibility, $\mu_\chi(L)$, to the following scaling law
\begin{equation}
    \mu_\chi(L) = \mu_c(\infty) + a/L + b/L^2,
    \label{scalingF}
\end{equation}
where $\mu_c(\infty)$, $a$, and $b$ are fitting parameters.

As depicted in Fig.~\ref{qmu2nn} (d), we find relation (\ref{scalingF}) to be
\begin{equation}
    \mu_\chi(L) = 6.66 - 104.546/L + 1648.85/L^2,
\end{equation}
from which wee see that, for $L > b/a \simeq 16$, the term proportional to $1/L$ dominates and the non-standard scaling discussed above takes place. From the same relation we obtain the critical chemical potential $\mu_c=6.66$ for $L\to \infty$.

\begin{figure}[!hbt] 
  \begin{center}
    \includegraphics[width=0.89\columnwidth]{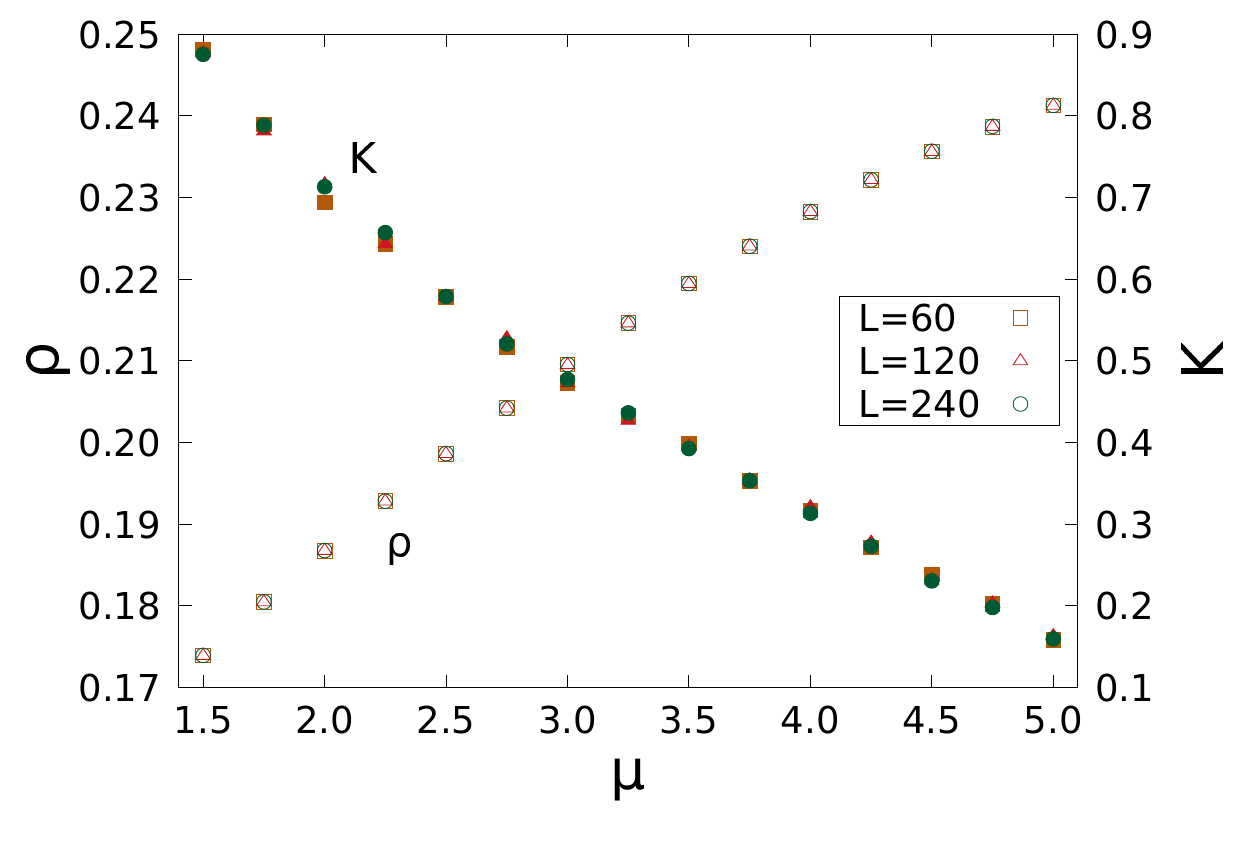}
    \caption{Density (empty symbols) and compressibility (solid symbols) as a function of $\mu<5.0$ for different lattice sizes in the 2NN case. In this regimen, we find no inflection point in density and no sign of phase transition in compressibility, even for large values of $L$.}
    \label{rhomu2nn}
  \end{center}
\end{figure}

Another interesting result is that even with snapshots in Fig.~\ref{snap2nn} (panels b and c) showing a domain-like configuration different from a fluid one, our simulations show no inflection point in density or any signs of phase transition in compressibility for $\mu<5.0$ (Fig.~\ref{rhomu2nn}).

In order to better understand the domain-like configurations, we define a local parameter $\psi_{6}(\vec{r})$ as the occupancy of the six sites corresponding to the nearest neighbors of order six (see Fig.~\ref{exclusion}) of a given particle at position $\vec{r}$ as
\begin{equation}
    \psi_{6}(\vec{r}) = \sigma_{\vec{r}}\sum_{\langle6NN\rangle}\sigma_i,
    \label{eq:psi6}
\end{equation}
where $\sigma_i=1$ if site $i$ is occupied and zero otherwise.

\begin{figure}[thb] 
  \begin{center}
    \includegraphics[width=.95\columnwidth]{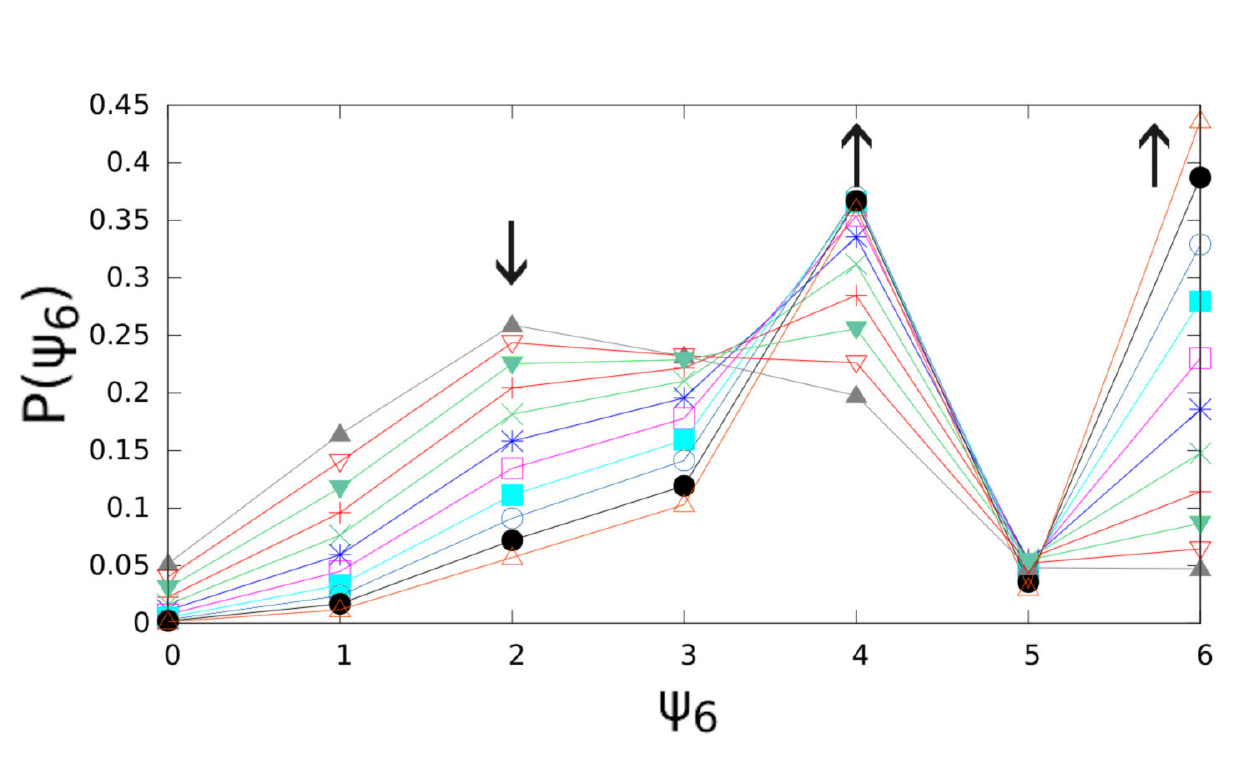}
    \caption{Probability distribution of $\psi_6$ (occupancy of neighbors of order six of a given particle) for $L=100$ in the 2NN case. Arrows show direction of increasing $\mu$, namely $3.6(\blacktriangle$), $3.8(\triangledown$), $4.0(\blacktriangledown$), $4.2(+)$, $4.4(\times)$, $4.6(*)$, $4.8(\square)$, $5.0(\blacksquare)$, $5.2(\circ)$, $5.4(\bullet)$ and $5.6(\triangle)$. Our results show that, for $\mu<3.6$, almost all particles have less than four neighbors of order six occupied ($\psi_6<4$), meaning the system is in a fluid-like configuration. As $\mu$ is increased, both domain-boundary ($\psi_6=4$) and domain-bulk ($\psi_6=6$) particles are predominant, indicating solid-like configurations.}
    \label{psiHist2nn}
  \end{center}
\end{figure}

With sublattices definitions as shown in Fig.~\ref{sl2nn}, neighbors of order six of a given particle are the nearest sites lying on the same sublattice as the particle itself. Therefore, it is straightforward to check that domain-bulk particles have all neighbors of order six occupied ($\psi_{6}=6$), whereas in domain-boundary particles, due to sliding freedom (see Fig.~\ref{ws2nn}), only four should be occupied ($\psi_{6}=4$). This local parameter allows us to investigate how domains grow as chemical potential is increased. Figure~\ref{psiHist2nn} shows the probability distribution of $\psi_{6}$ for different $\mu$. As can be seen, for $\mu\sim3.6$ almost no domain-bulk or domain-boundary particles exist, indicating fluid-like configurations.  As $\mu$ is increased, the system becomes more and more solid-like, until the phase transition related to order parameter $Q_2$ (Eq. \ref{eq:q2nn}) occurs and translational symmetry is restored along two of the three lattice directions.

\begin{figure}[thb] 
  \begin{center}
    \includegraphics[width=6cm]{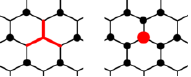}
    \caption{Equivalence of exclusion area in a system of a $Y$-shaped particles with no superposition (left) and the 2NN hardcore model (right).} 
    \label{yshap}
  \end{center}
\end{figure}

We also notice the effective area of a particle in the case $k=2$ is the same as a $Y$-shaped particle on the honeycomb lattice (see Fig.~\ref{yshap}). In~\cite{yShapedRaj}, authors hint at a possible columnar phase as a second order perturbation in a full-packed system of $Y$-shaped particles on the honeycomb lattice. In their brief discussion, the following scenario is presented: starting in a solid-like phase at maximum density, first a transition to a columnar phase takes place as density is decreased. Second, a transition to another solid-like phase followed by a disordered one, or a direct transition to a disordered phase, should happen. Here we provided numerical evidence that the columnar phase undergoes a phase transition to a solid-like phase that decays, as chemical potential is decreased, into fluid-like configurations without any signs of a second phase transition.

In summary, the 2NN case undergoes a two step melting as chemical potential is decreased. At densities close to full packing the system breaks into independent slabs of size $2\times L$ with sliding freedom along one of the three lattice directions. As density is decreased, a phase transition occurs and all four sublattices become equally occupied. In these configurations, several ordered, solid-like domains, are observed. Further decrease in density shrinks these domains until, without any signs of a phase transition, symmetry along all three directions is restored and the system reaches fluid-like configurations.

\subsection{\label{subsec:3nn}Up to third neighbors exclusion ($k=3$)}

\begin{figure}[thb] 
  \begin{center}
    \includegraphics[width=0.62\columnwidth]{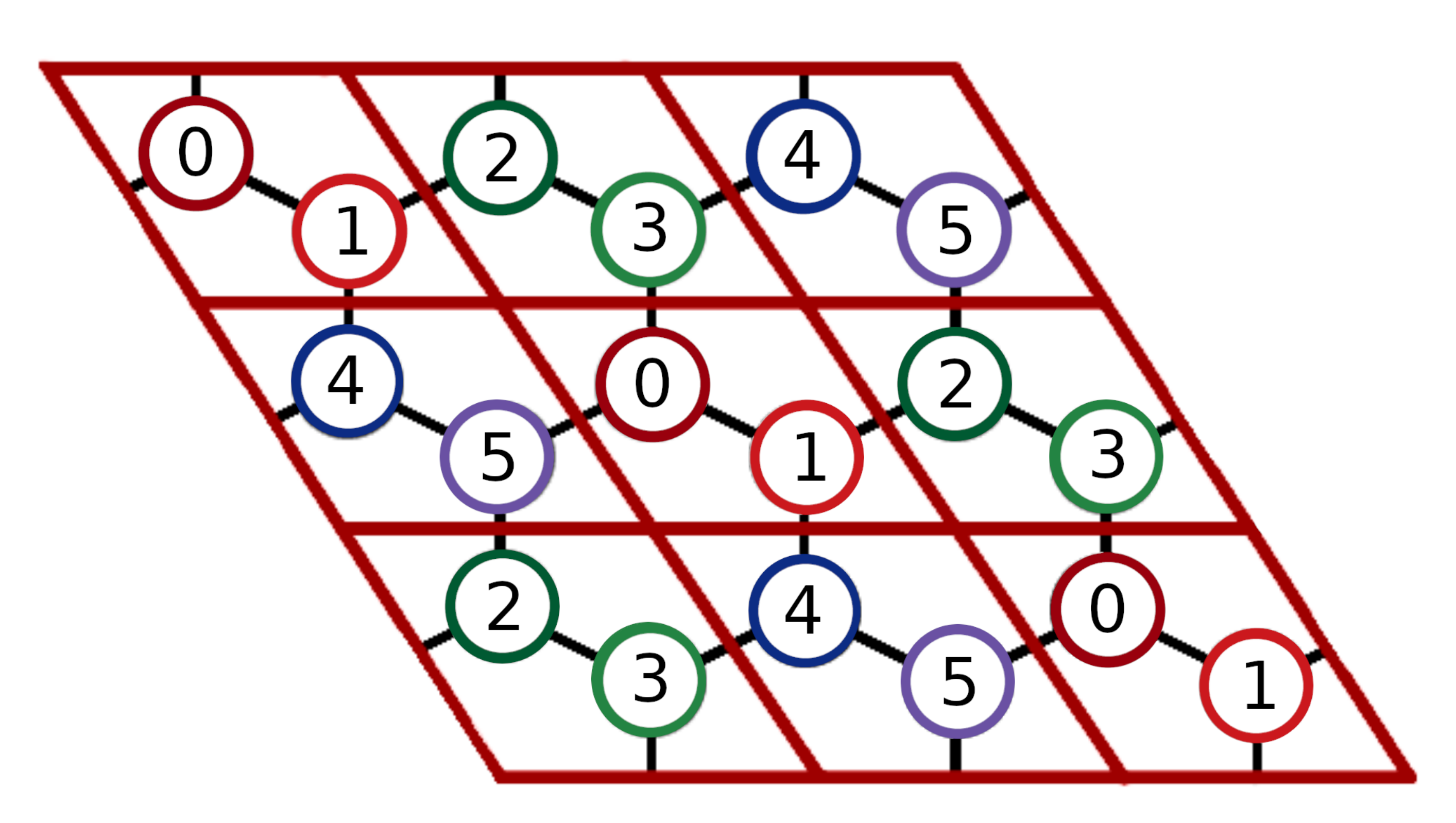}
    \includegraphics[width=0.36\columnwidth]{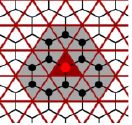}
    \caption{Sublattice definitions (left) and (right) equivalence between the $k=3$ case and a model of triangular trimers on the triangular lattice where no site may be occupied by more than one trimer. The excluded area of a 3NN particle on the honeycomb lattice (black dots) is the same as the excluded area of a triangular trimer on the triangular lattice (shaded triangular faces).}
    \label{trimers}
  \end{center}
\end{figure}

One interesting feature of the 3NN model is that it can be mapped onto a system of triangular trimers on a triangular lattice where each site may be occupied by only one trimer (right panel of Fig.~\ref{trimers}). The model of triangular trimers on the triangular lattice at full packing was studied in~\cite{verberkmoesTriang} and, within a 2D subset of the 4D parameter space, the authors found a phase transition related to symmetry break of up and down trimers. In this 2D subset, it is shown that the three sublattices of up (down) trimers are equally occupied.

In the 3NN model on the honeycomb lattice, this up/down trimers symmetry breaking is related to symmetry breaking in the occupancy of $A$ and $B$ sites. Since we are using 
the grand canonical (and not the canonical) ensemble, their assumptions to solve the 2D parameter space are not expected to be satisfied, except for some very unlikely configurations. 

\begin{figure}[thb] 
  \begin{center}
    \includegraphics[width=0.99\columnwidth]{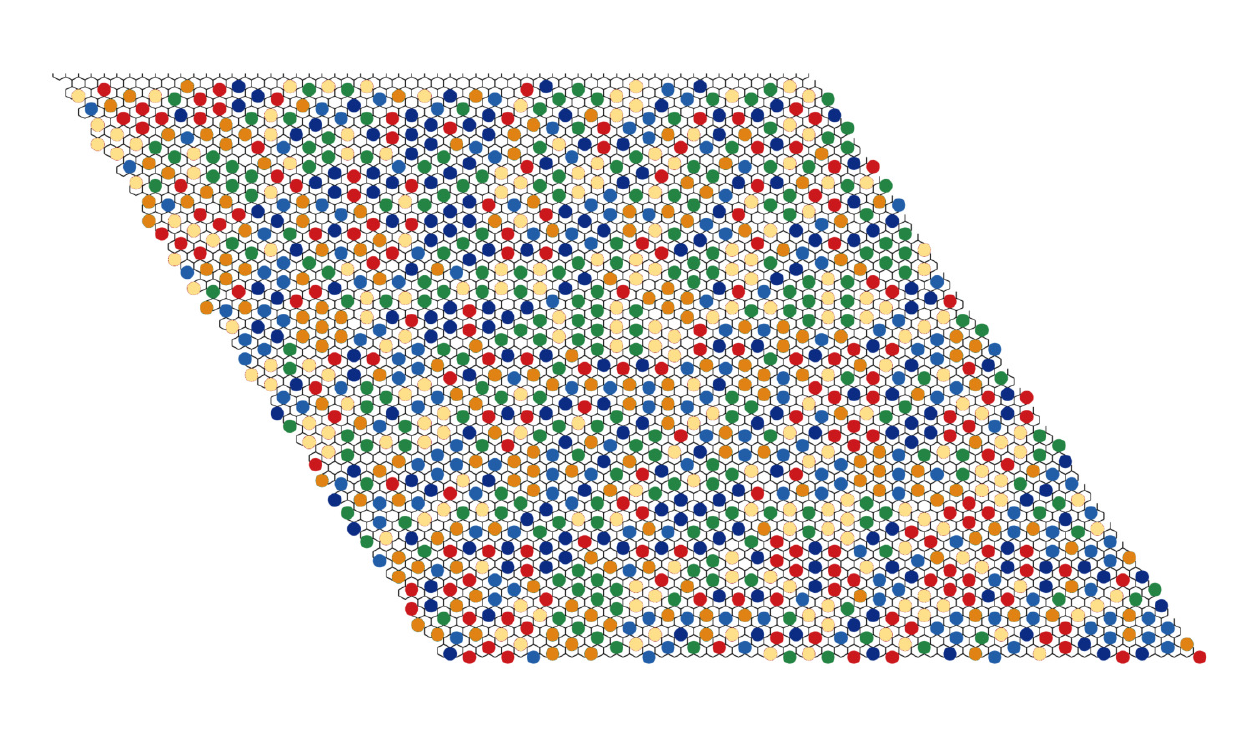}
    \caption{A typical configuration for $L=60$ at $\mu[\rho]=9.0[0.1638]$ for the 3NN case. The packing fraction is $\phi=0.9825$. It is possible to see several different forms of local, short range ordering, but no global ordering is reached within our simulations.}
    \label{conf3nn}
  \end{center}
\end{figure}

Defining sublattices in an equivalent way to the ones in the aforementioned reference (Fig~\ref{trimers}, left panel), we find no phase transition as density is increased. Moreover, the full packing configurations ($\rho_{max}=1/6$) do not show any symmetry breaking in sublattices occupation or in the occupancy of $A$ or $B$ sites. A typical configuration at very high densities can be seen in Figure~\ref{conf3nn}.

\begin{figure}[thb] 
  \begin{center}
    \includegraphics[width=.98\columnwidth]{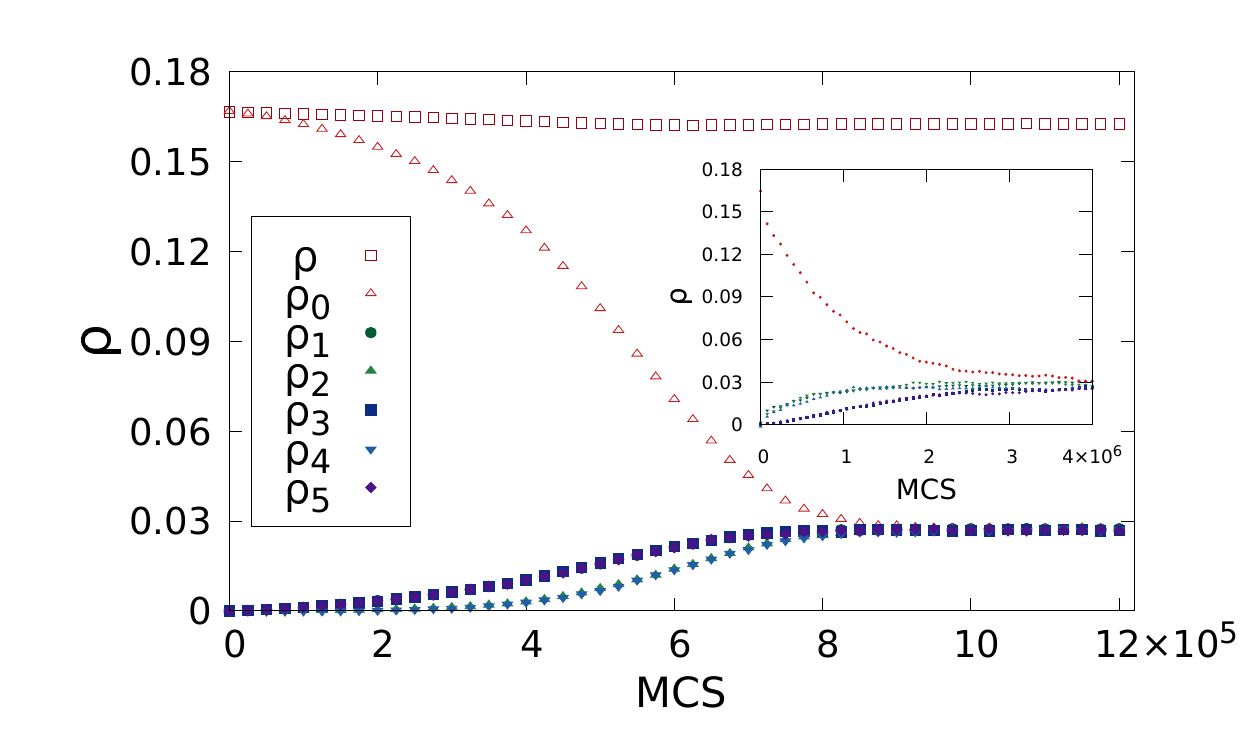}
    \caption{Time series for the 3NN case with $L=600$ and $\mu=8.0$ showing how the system relaxes into configurations with all sublattices evenly occupied when an initial condition with only one sublattice is chosen. Inset: canonical simulation for $L=120$ with $N=N_{max}-L/2$ ($\phi=0.9875$) particles starting in one sublattice. The system quickly reaches configurations with all sublattices equally occupied.}
    \label{serie3nn}
  \end{center}
\end{figure}

Figure~\ref{serie3nn} shows how the system relaxes into configurations where particles are evenly distributed along all sublattices for $\mu=8.0$ and $L=600$, even when a full packing configuration in one sublattice is chosen as initial condition. In the same figure, a second panel shows how a canonical simulation at fixed density $\rho\simeq0.16458$ ($\phi=0.9875$) evolves into a state where all sublattices are equally occupied.

\begin{figure}[thb] 
  \begin{center}
    \includegraphics[width=.44\columnwidth]{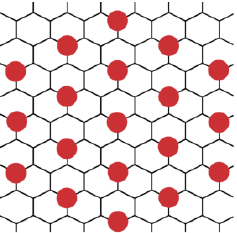}
    \hspace{.3cm}
    \includegraphics[width=.44\columnwidth]{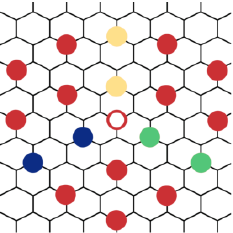}
    \caption{One possible fully packed configuration for the 3NN case where only one sublattice is occupied (left) and (right) how the removal of one particle (empty circle) creates a sliding instability in three of the five remaining sublattices, represented by different colors (shades of gray). This sliding instability allows the system to reach disordered configurations even at high densities, corroborating the lack of global ordering observed in our simulations.}
    \label{3nnInstability}
  \end{center}
\end{figure}

To further support the lack of phase transition at packing fractions up to $\phi=0.9875$, we show in Fig.~\ref{3nnInstability} how the removal of one particle creates instability in three of the five remaining sublattices, giving rise to an $Y$-shaped domain boundary.

Since the instability is in all three lattice directions, the argument presented in Ref.~\cite{yShapedRaj} for the triangular lattice predicts no columnar phase. Therefore, from a full-packed configuration the system decays directly into a disordered one and the observed lack of phase transition in our simulations at densities below $\rho_{max}$ is in accordance with their arguments.

\subsection{Up to fourth neighbors exclusion ($k=4$)}
\label{subsec:4nn}
We expect this model to undergo a symmetry break in occupancy of $A$ and $B$ sites at full packing (see Sec.~\ref{sec:conjec} for discussion). A particle of type $A$ has (on the $A$ lattice) the same exclusion as the 1NN hardcore model on the triangular lattice (see. Fig~\ref{redeTriang}) exactly solved by Baxter~\cite{baxter_hh}, which has a phase transition at $\mu=2.406$ on the 3-state Potts model universality class. Therefore, an interesting scenario appears. On one hand, if the $A$-$B$ symmetry break occurs for $\mu_{AB}<2.406$, we should see at least two phase transitions, the first being in the occupancy of $A$ and $B$ sites and the second a transition to a solid phase. On the other hand, if $\mu_{AB}>2.406$, only one phase transition should be expected.

Since presence of another type of particle on the honeycomb lattice at high densities, as compared to its triangular counterpart, can be interpreted as presence of impurities, which generally increase the critical chemical potential, we expect the second case ($\mu_{AB}>2.406$) to be true.

\begin{figure*}[!htb] 
  \begin{center}
    \includegraphics[width=0.65\columnwidth]{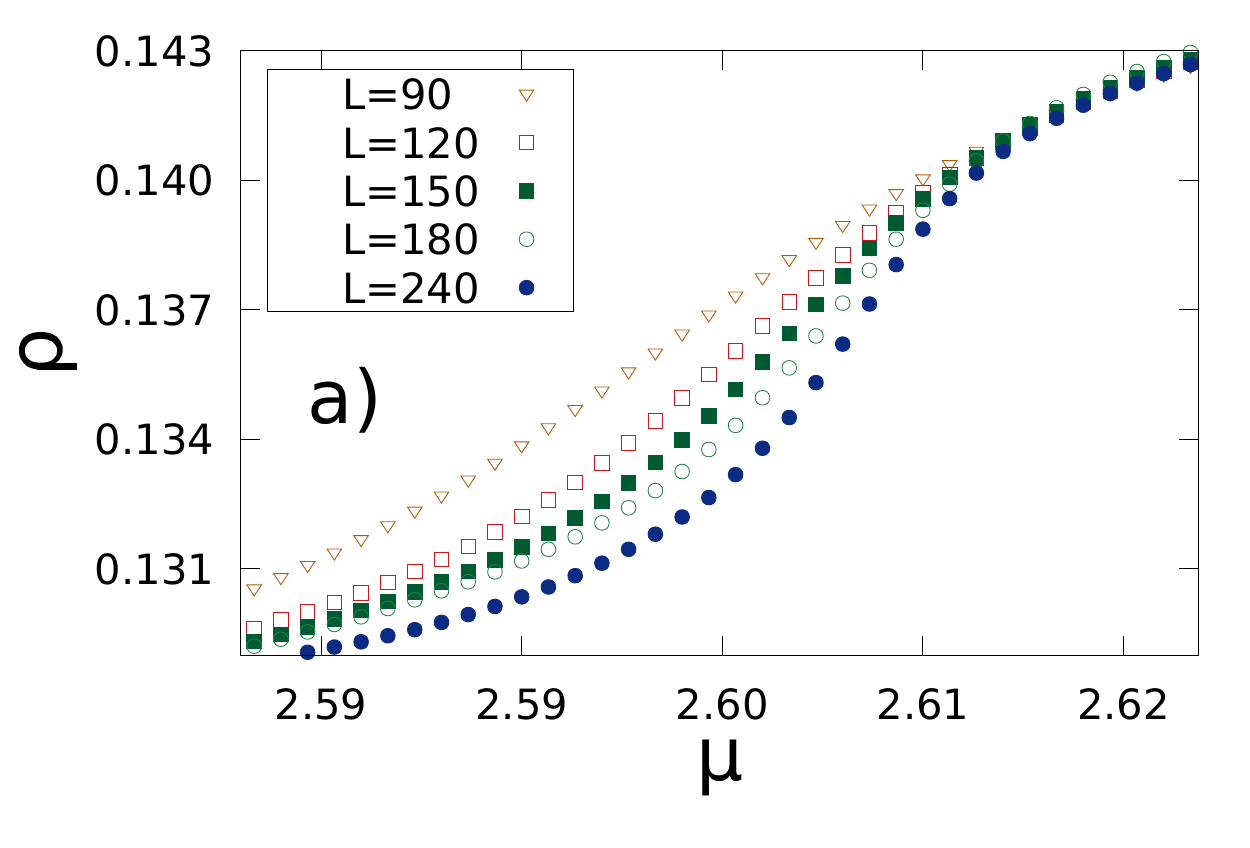}
    \includegraphics[width=0.65\columnwidth]{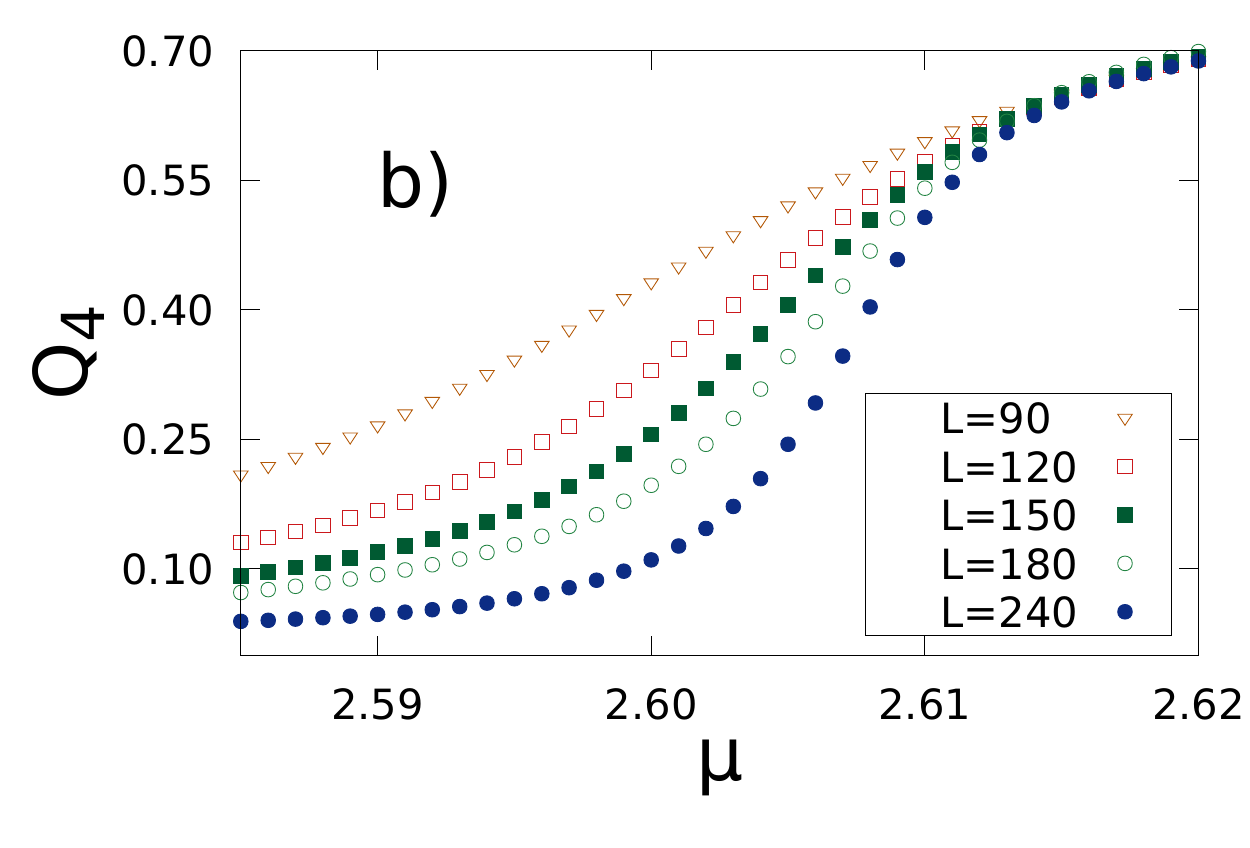}
    \includegraphics[width=0.65\columnwidth]{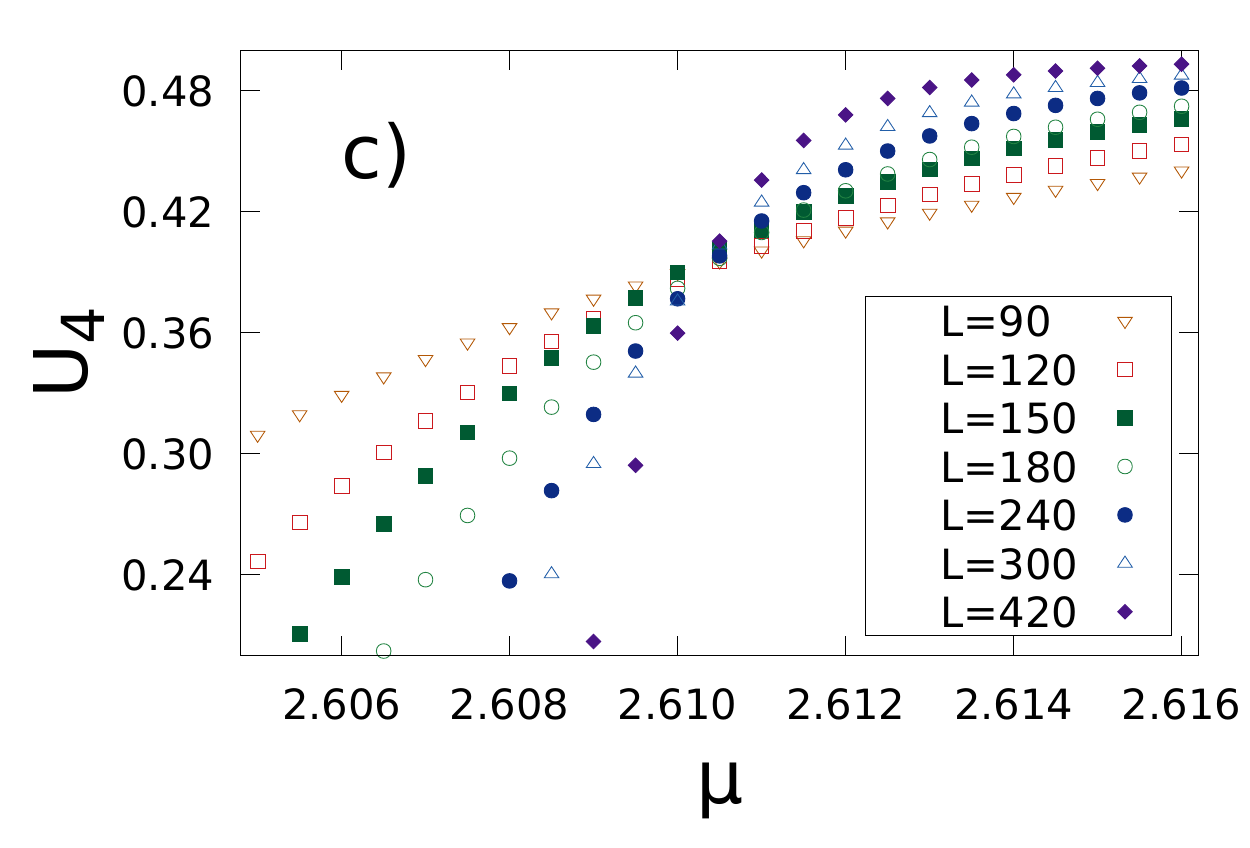}
    \includegraphics[width=0.65\columnwidth]{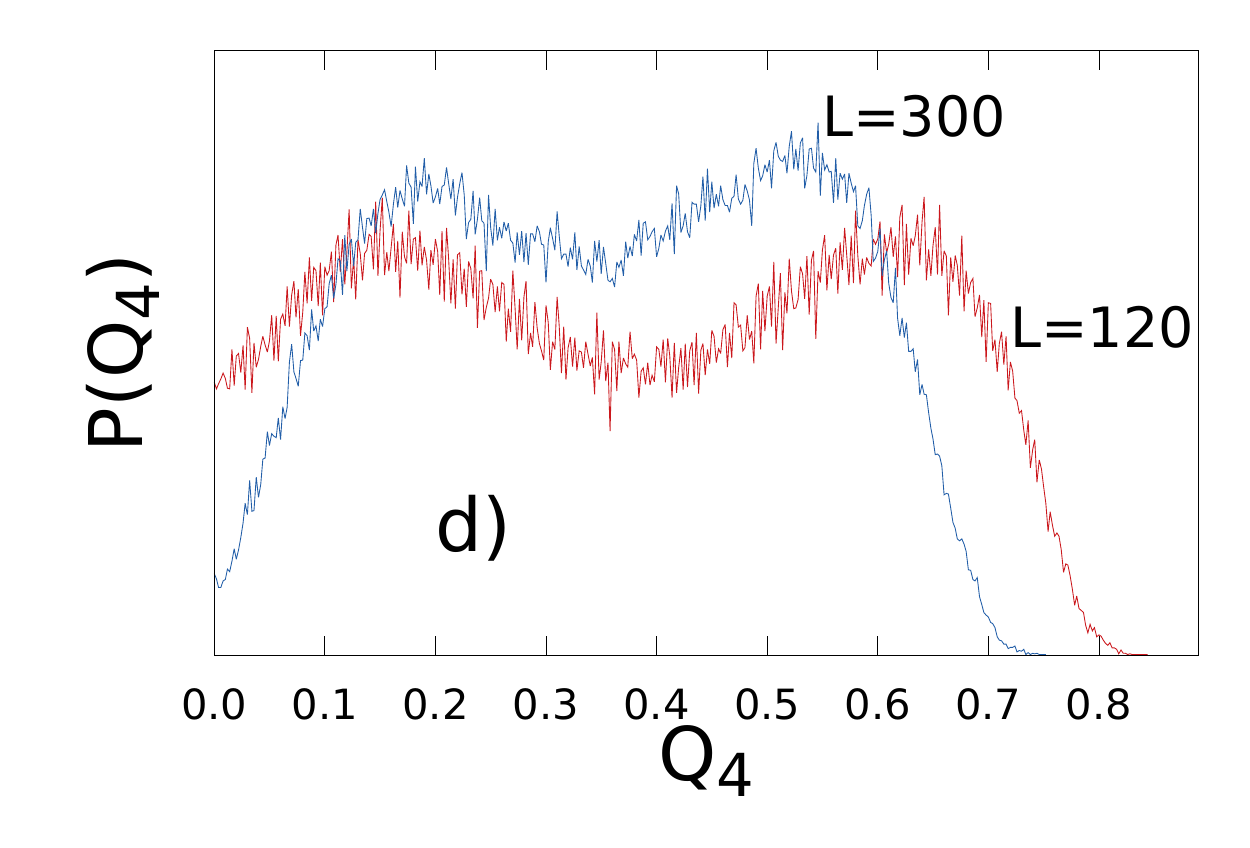}
    \includegraphics[width=0.65\columnwidth]{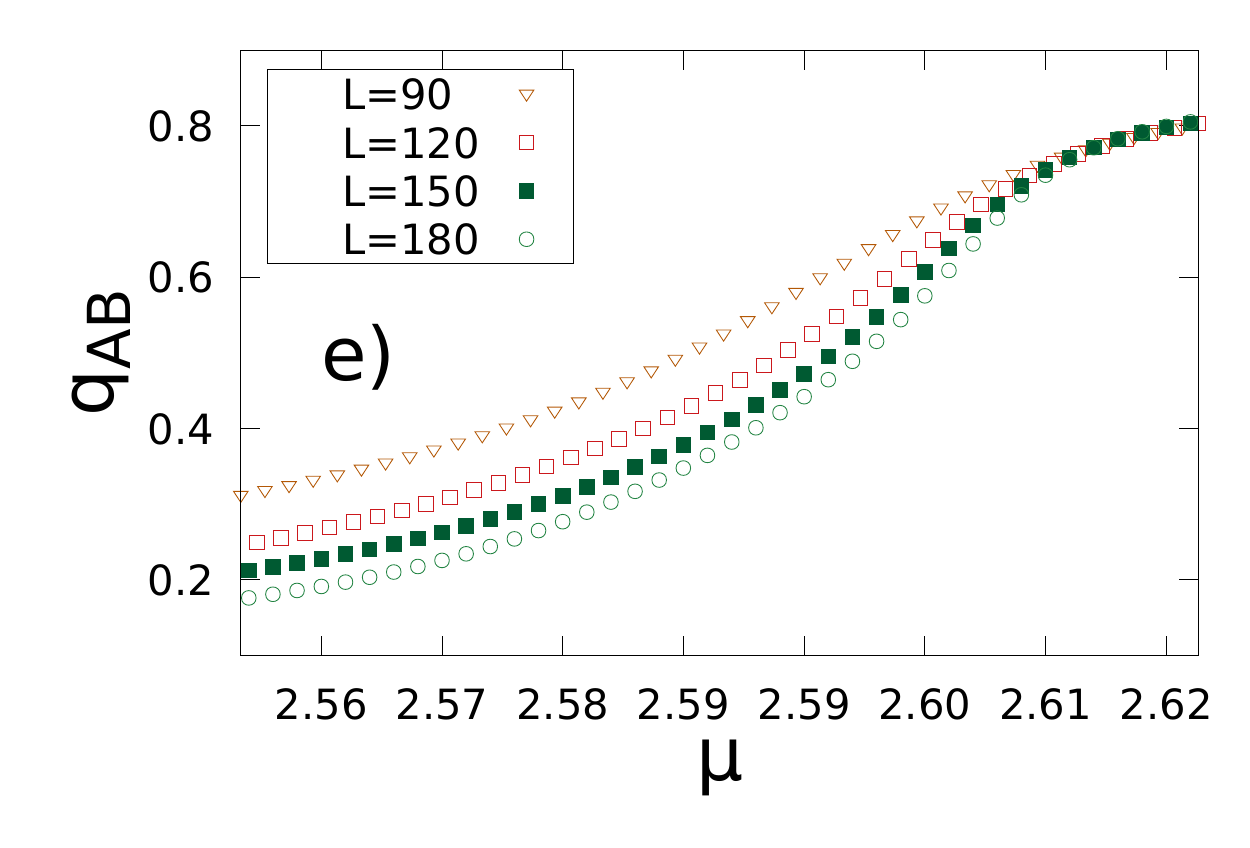}
    \includegraphics[width=0.65\columnwidth]{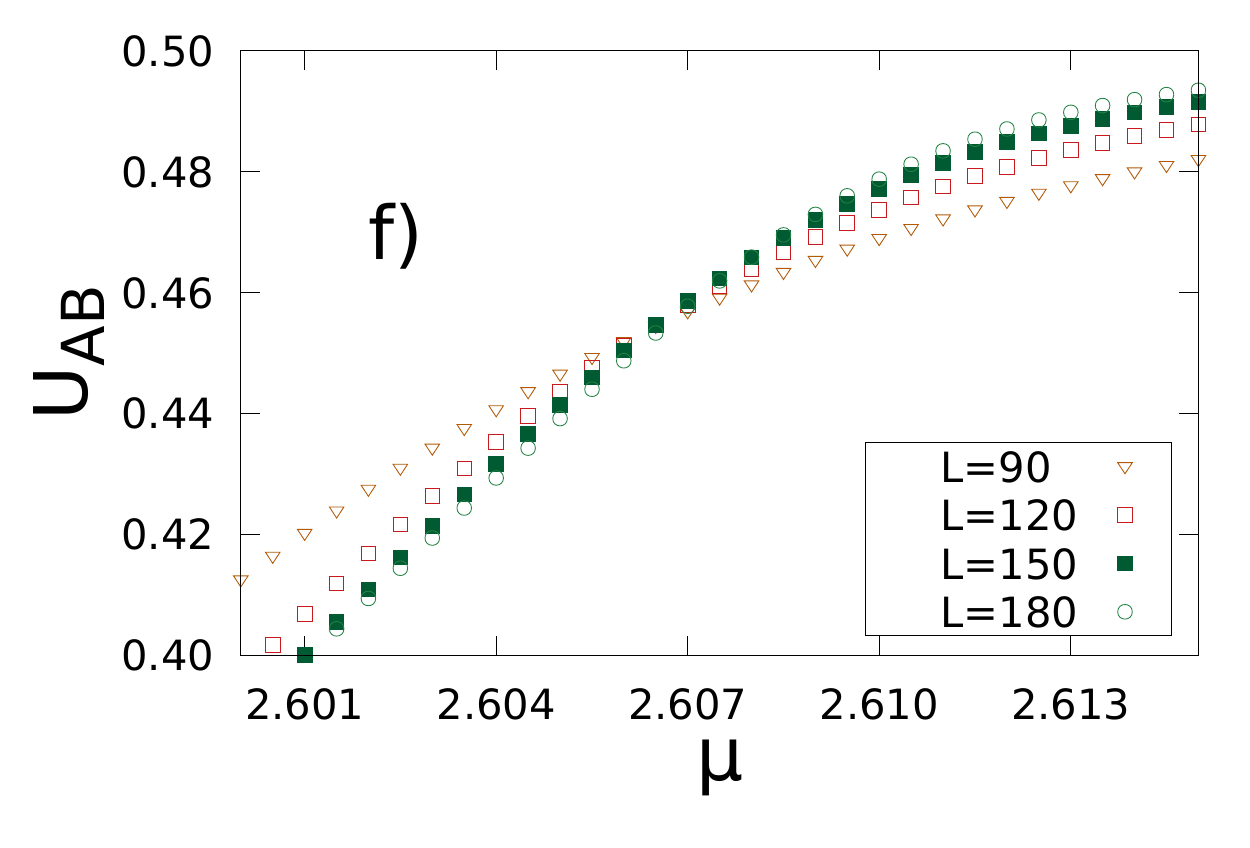}
    
    \caption{Results for the 4NN case. From top left to bottom right: (a) Density, (b) order parameter $Q_4$ and (c) its Binder cumulant as function of $\mu$. Panels (e) and (f): same as (b) and (c) for $q_{AB}$. The intersection points in (c) and (f) are $\mu_{Q_4}=2.6105$ and $\mu_{q_{AB}}=2.607$, respectively. 
    Panel (d) shows the histograms of $Q_4$ for two different $L$ near the phase transition. Although there are two peaks, they get closer with increasing system size and should eventually merge. This, together with the finite size scaling analysis, characterizes the transition as being of second order.}
    \label{results4nn}
  \end{center}
\end{figure*}

To characterize phases, we define the following quantities:
\begin{equation}
  \label{pos4nn}
  \begin{split}
    q_A & = | \rho_1 - \rho_3 | + | \rho_1 - \rho_5 | + |\rho_3-\rho_5|, \\
    q_B & = | \rho_0 - \rho_2 | + | \rho_0 - \rho_4 | + |\rho_2-\rho_4|,\\
  \end{split}
\end{equation}
with sublattices as in the 3NN case (Fig.~\ref{trimers}). Quantities in equations (\ref{pos4nn}) are the same defined on the 1NN model on the triangular lattice, one for each type of site. We define two order parameters as
\begin{equation}
  \begin{split}
    q_{AB} & = 6|\rho_A - \rho_B|, \\
    Q_4 &= 3|q_A - q_B|,\\
  \end{split}
\end{equation}
with $\rho_A$ ($\rho_B$) being the sum over odd (even) sublattice densities and measure their Binder cumulant given by

\begin{equation}
    U_Q = 1 - \frac{\langle Q^4\rangle}{2\langle Q^2\rangle^2}.
\end{equation}

\begin{figure}[!hbt]
    \begin{center}
        \includegraphics[width=0.99\columnwidth]{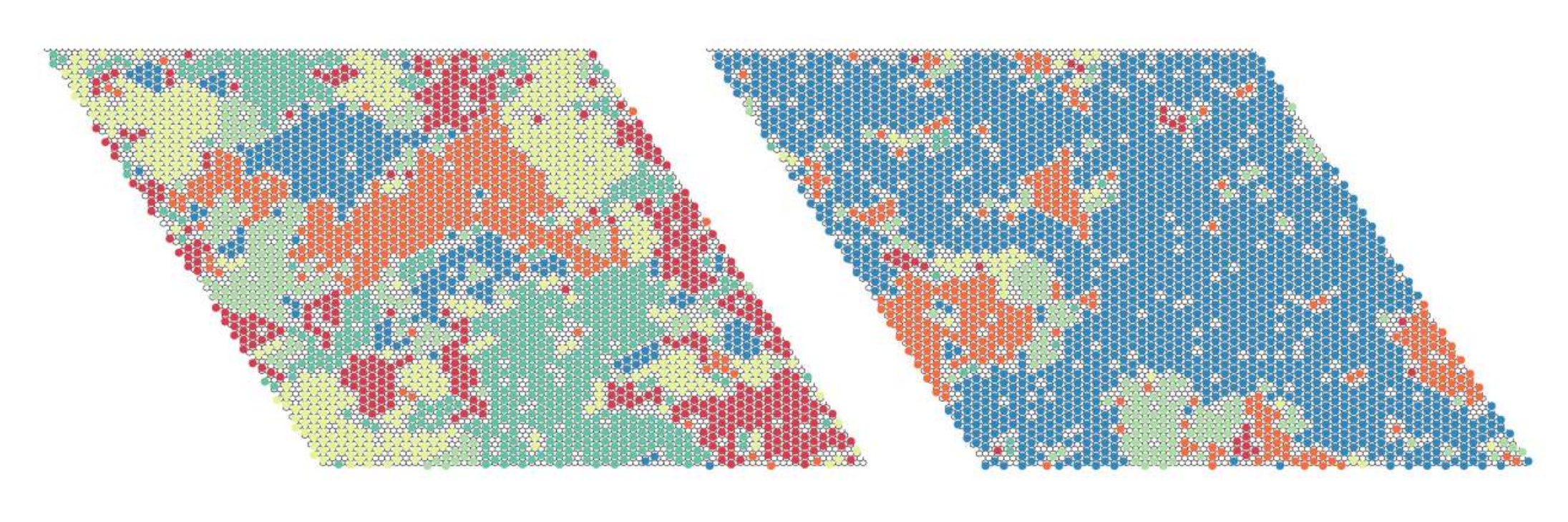}
        \caption{Snapshot ($L=60$) of two configurations for the 4NN case with $\rho=0.131$ (left) and $\rho=0.139$ (right) near the phase transition ($\mu=2.6$). The left panel shows all six sublattices (different colors) equally occupied while, on the right panel, one sublattice dominates. The right panel also shows the broken $A$-$B$ symmetry since there are mainly particles in one type of site.}
        \label{snap4nn}
    \end{center}
\end{figure}

The curves of density and order parameters $Q_4$ and $q_{AB}$ are presented in Fig.~\ref{results4nn}. Our results for the intersection point of the Binder cumulant show an $A$-$B$ phase transition at $\mu=2.607$ (panel (f)) and a sublattice transition  at $\mu=2.6105$ (panel (c)). While it is tempting to conclude these transitions to be two different critical points, numerical precision does not allow us to do so. Moreover, it is not completely clear whether there are any significant differences in particles arrangement between the fluid ($\mu<2.607$) and the (possible) intermediary ($2.607<\mu<2.6105$) phase. Since our simulations do not allow us to distinguish between these phases, we regard the two transitions to be the same. The snapshots in Fig.~\ref{snap4nn} show typical configurations at $\mu$ near the phase transition.

Next, it is possible to see two different peaks in histograms of order parameter $Q_4$ in panel (d) of Fig.~\ref{results4nn} which clearly get closer with increasing $L$ and should merge in the thermodynamic limit. This behavior is expected only in continuous transitions since the peaks should separate with increasing $L$ if the transition was of first order nature.

Through finite size scaling analysis, we find a good collapse of curves for $\mu_c=2.6108$ with a set of critical exponents $\gamma=1.28$, $\nu=0.83$ ($\gamma/\nu\simeq 1.542$) and $\beta=0.1$, which are very close to the $3$-state Potts model exponents, except for $\gamma_{q=3}=13/9\simeq 1.44$. These exponents corroborate the continuous nature of the transition.

\begin{figure}[!hbt]
    \begin{center}
        \includegraphics[width=0.8\columnwidth]{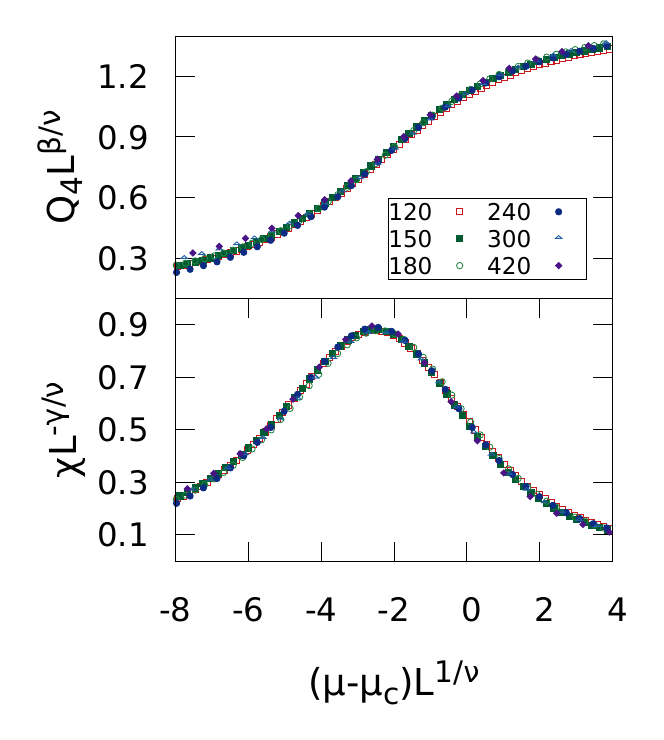}
        \caption{Curves of the 4NN order parameter $Q_4$ and its susceptibility collapse after re-scaling with critical exponents $\gamma=1.28$, $\nu=0.83$ and $\beta=0.1$, which are close to the 3-state Potts model exponents, except for $\gamma$. 
        }
        \label{col4nn}
    \end{center}
\end{figure}

In Fig.~\ref{col4nn} we show the collapsed curves of order parameter $Q_4$ and its susceptibility after re-scaling with the critical exponents found.

\subsection{\label{subsec:5nn}Up to fifth neighbors exclusion ($k=5$)}
Similar to the 4NN model, the full packing configuration of the case with up to fifth neighbors exclusion allows only one type of particle. This means that we expect an $A$-$B$ transition at high densities. After this transition, this model becomes similar to the 2NN case on the triangular lattice, studied in~\cite{ZhangDeng,tensorTriangular}, which has a phase transition at $\mu=1.75$.

\begin{figure}[thb]
    \begin{center}
        \includegraphics[width=0.7\columnwidth]{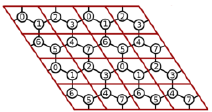}
        \caption{Sublattice definitions for the 5NN model.}
        \label{sl5nn}
    \end{center}
\end{figure}

We define sublattices as in Fig.~\ref{sl5nn} and measure the following order parameter
\begin{equation}
  Q_5=\frac{8}{7}\sum_{i=0}^7\sum_{j>i}^7|\rho_i - \rho_j|,
\end{equation}
which is equal to unity whenever only one sublattice is fully occupied. Since only one sublattice is occupied in the full packing configurations, the maximum density is $\rho_{max}=1/8$.

\begin{figure}[hbt] 
  \begin{center}
    \includegraphics[width=.95\columnwidth]{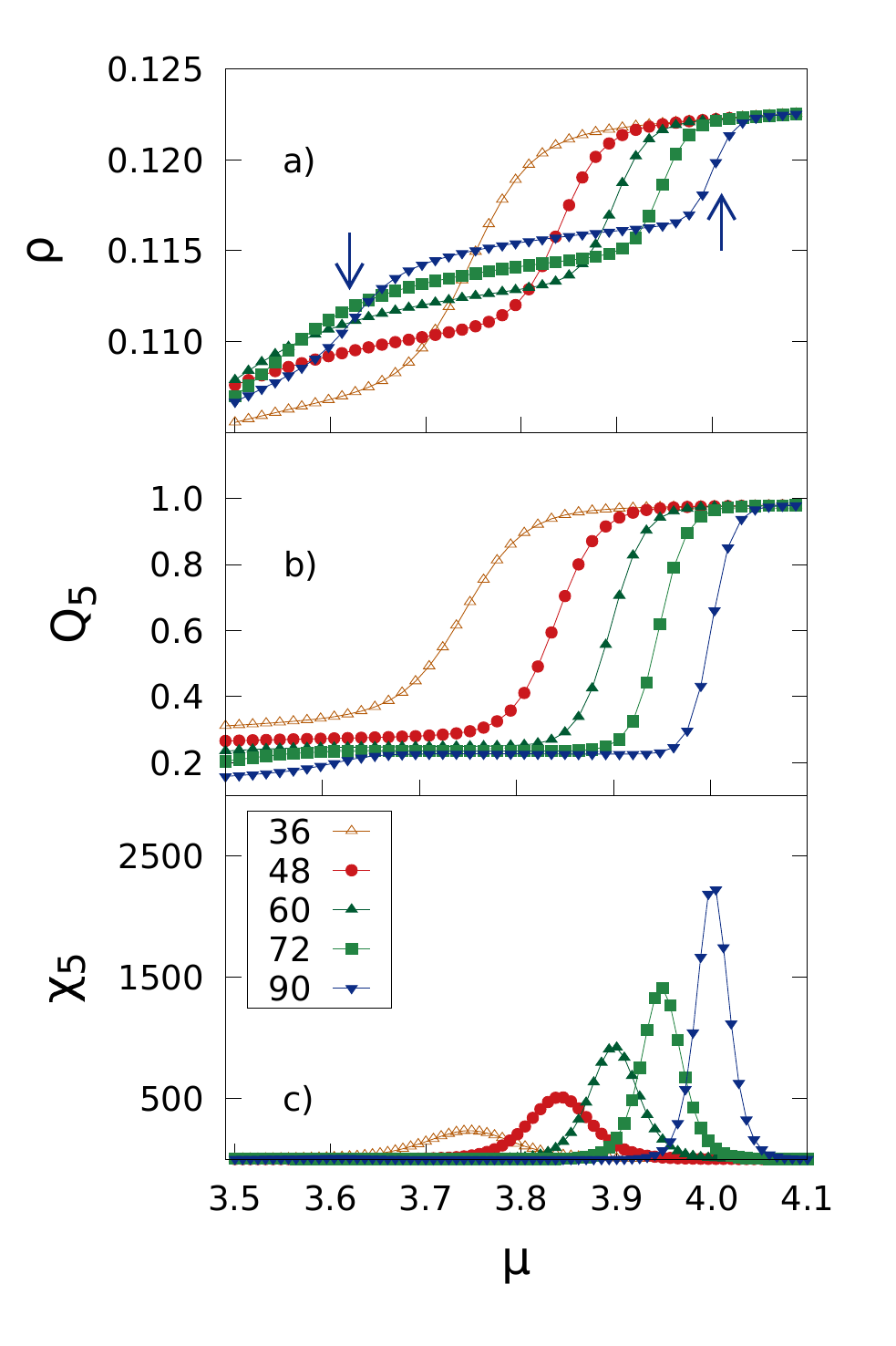}
    \caption{(a) Density $\rho$ for different lattice sizes as a function of chemical potential for the 5NN case showing the complex landscape of phase transitions that this model undergoes. The arrows indicate the two different inflection points for $L=90$, 
    indicating two separate phase transitions. Panel (b) shows the order parameter $Q_5$ and (c) its susceptibility. Results were obtained using the Wang-Landau sampling with adaptive windows.}
    \label{qmu5nn}
  \end{center}
\end{figure}

Different from the previous models, we find two transition points as indicated by the two inflection points in Fig.~\ref{qmu5nn} (a) and (b). 

In the first phase transition, i.e. at lower $\mu$, the system changes from a fluid-like phase into a domain-like phase. This domain phase is characterized by clusters with domain boundaries running along the entire length of all three lattice directions.

\begin{figure}[hbt]
  \begin{center}
   \includegraphics[width=\columnwidth]{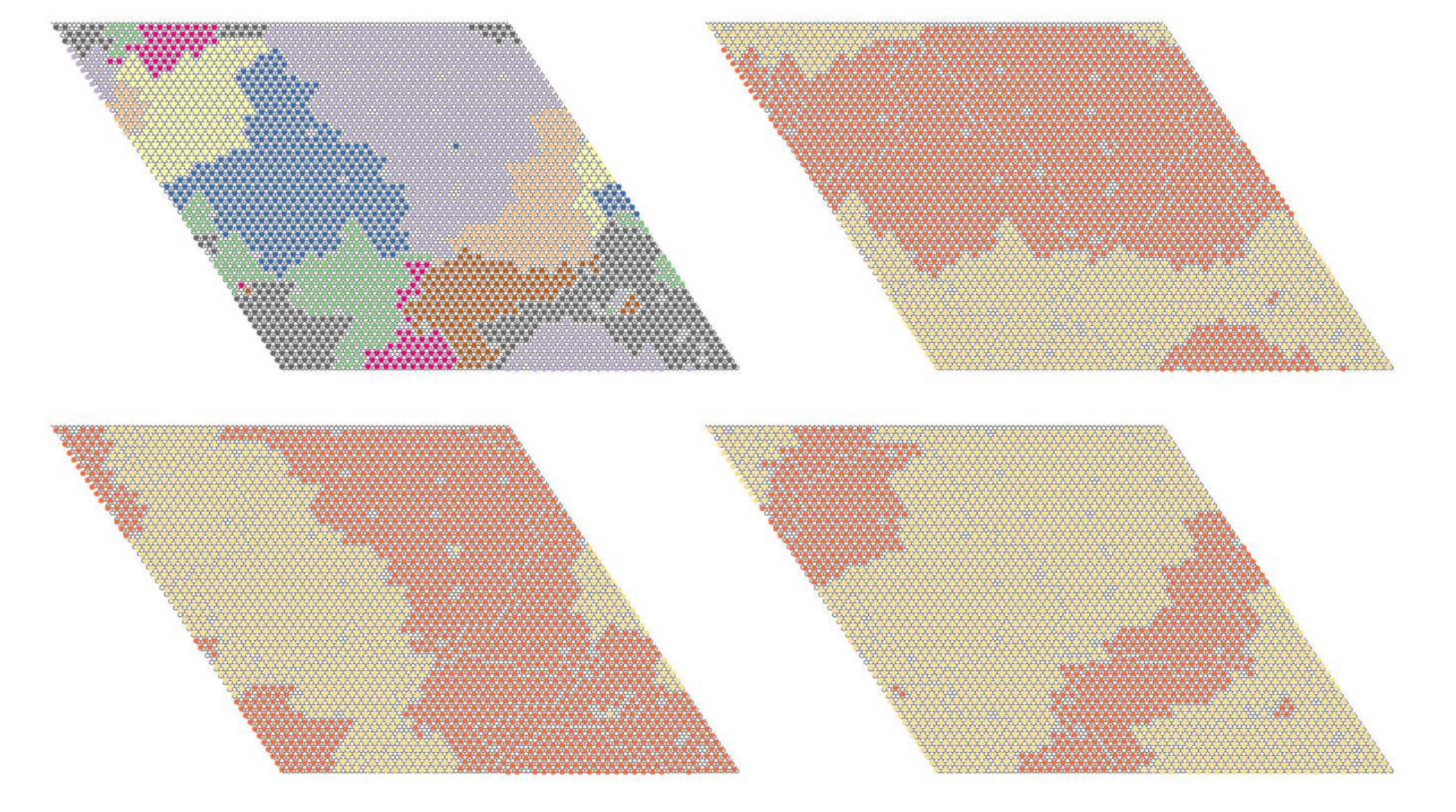}
    \caption{One single snapshot for the 5NN model at $\mu=3.9$. The top left panel shows all eight sublattices (different colors) and, in subsequent panels, the two color scheme represents the groups in Table~\ref{groups5nn}, showing how the system organizes into domains along the three directions. $L=102$ and $\rho=0.117$.}
    \label{snap5nn}
  \end{center}
\end{figure}

By grouping sublattices into sets of four as shown in Table~\ref{groups5nn}, and looking at snapshots (Fig.~\ref{snap5nn}) after the first transition at $\mu\simeq 3.9$, we clearly see how the system organizes into clusters along the three directions.

\begin{table}[h]
  \begin{center}
    \begin{tabular}{  c || c | c | c }
    
      $\theta$  & 0 & $\pi/3$ & $-\pi/3$\\ \hline
      Sublattices & \{0,1,2,3\} & \{0,1,5,6\} & \{1,2,6,7\}\\
       & and & and & and\\
       & \{4,5,6,7\} & \{2,3,4,7\} & \{0,3,4,5\}\\
    \end{tabular}    
    \caption{Groups of sublattices for the three lattice directions in the 5NN case. In the domain-like phases, these groups form clusters along the given direction (see snapshots in Fig.~\ref{snap5nn}).}
    \label{groups5nn}
  \end{center}
\end{table}

\begin{figure*}[thb] 
  \begin{center}
    \includegraphics[width=.66\columnwidth]{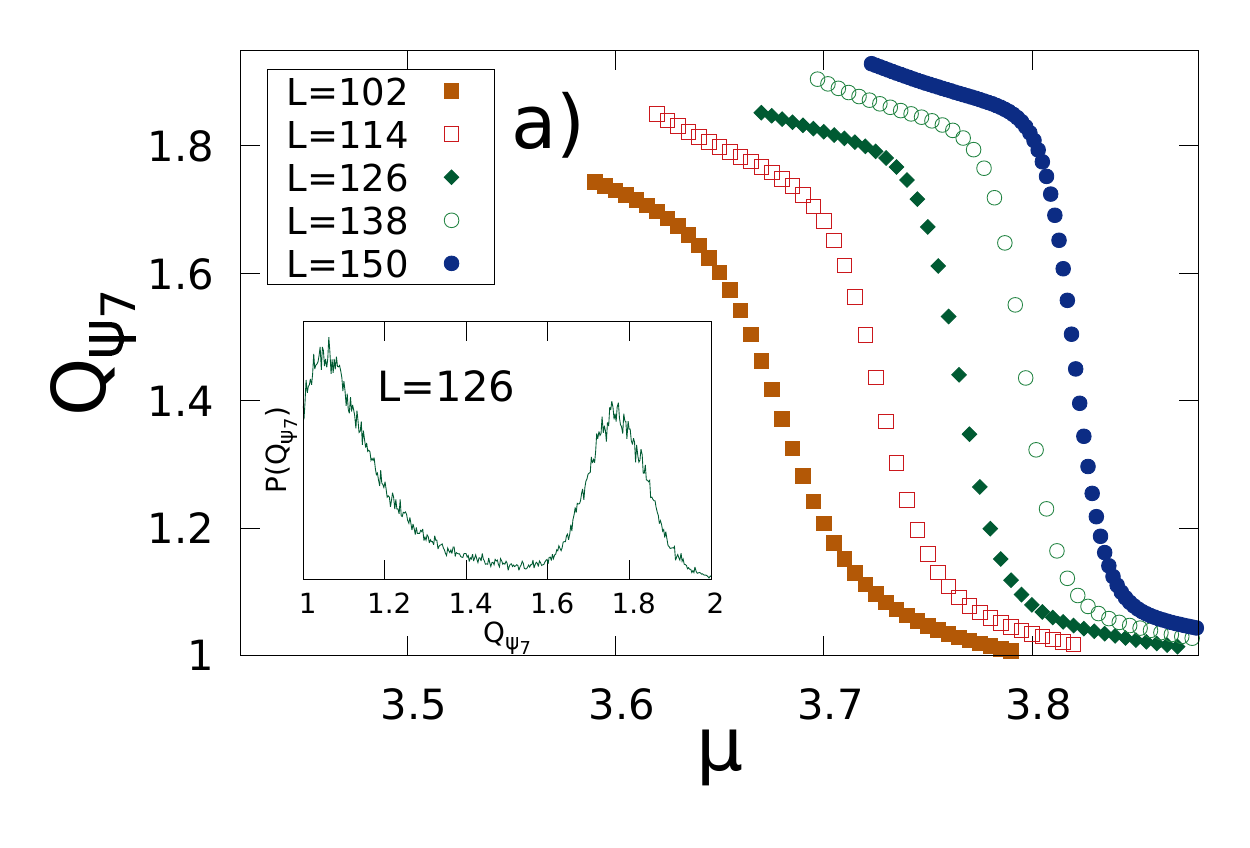}
    \includegraphics[width=.66\columnwidth]{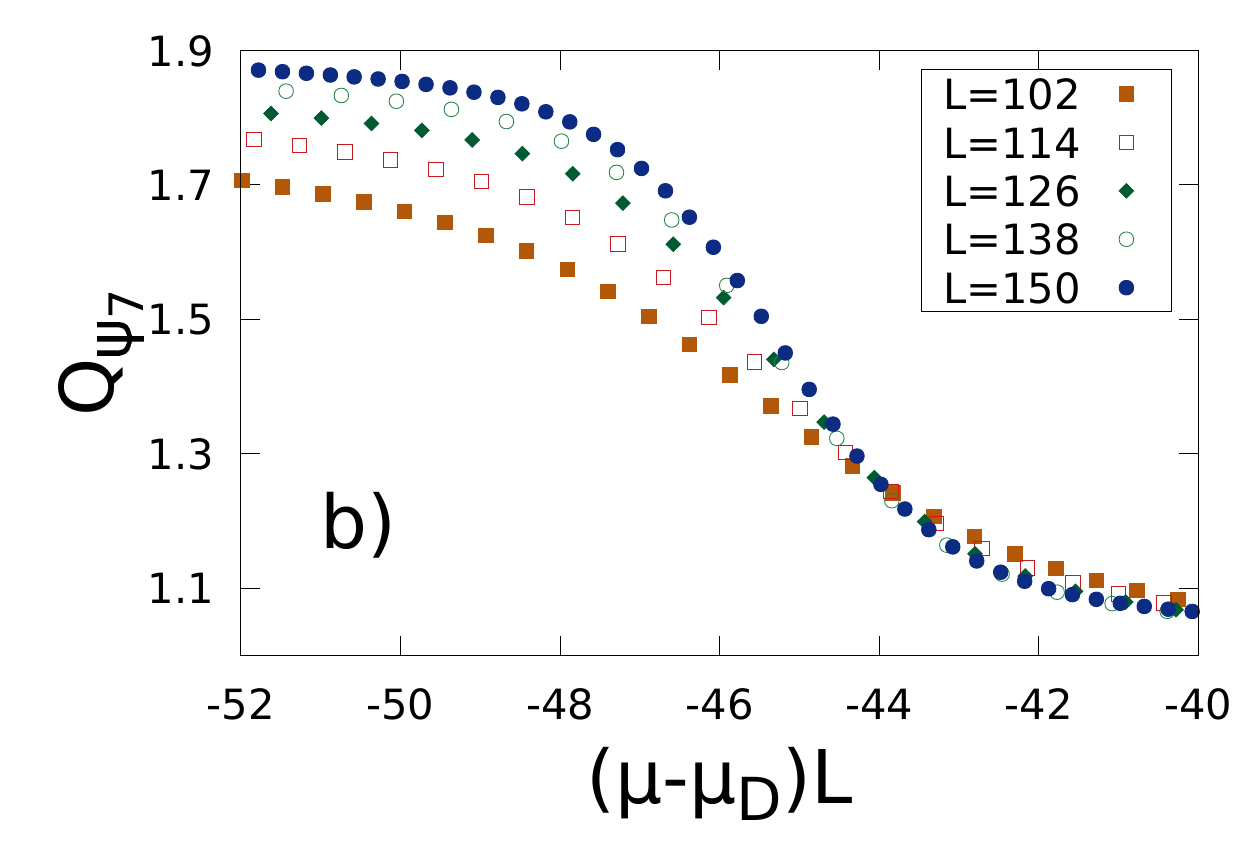}
    \includegraphics[width=.66\columnwidth]{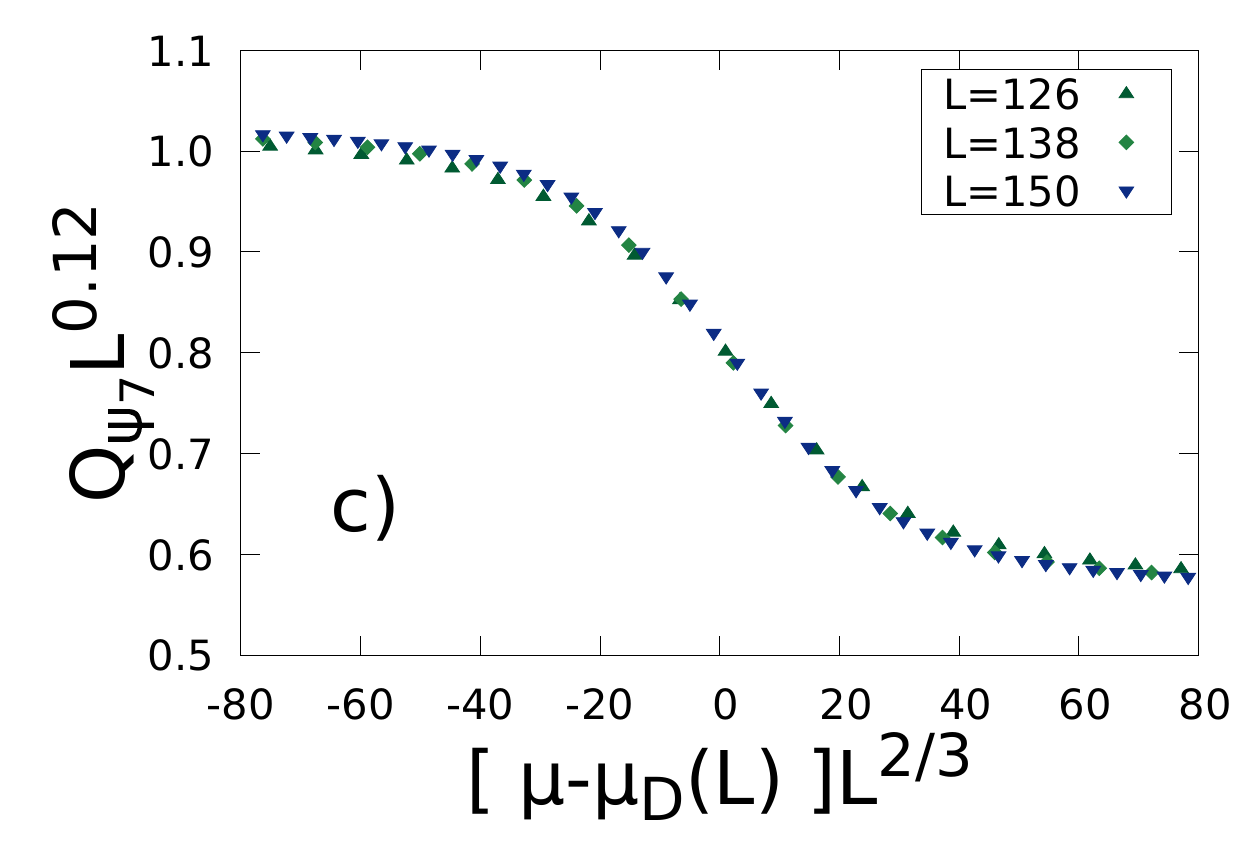}
    \includegraphics[width=.66\columnwidth]{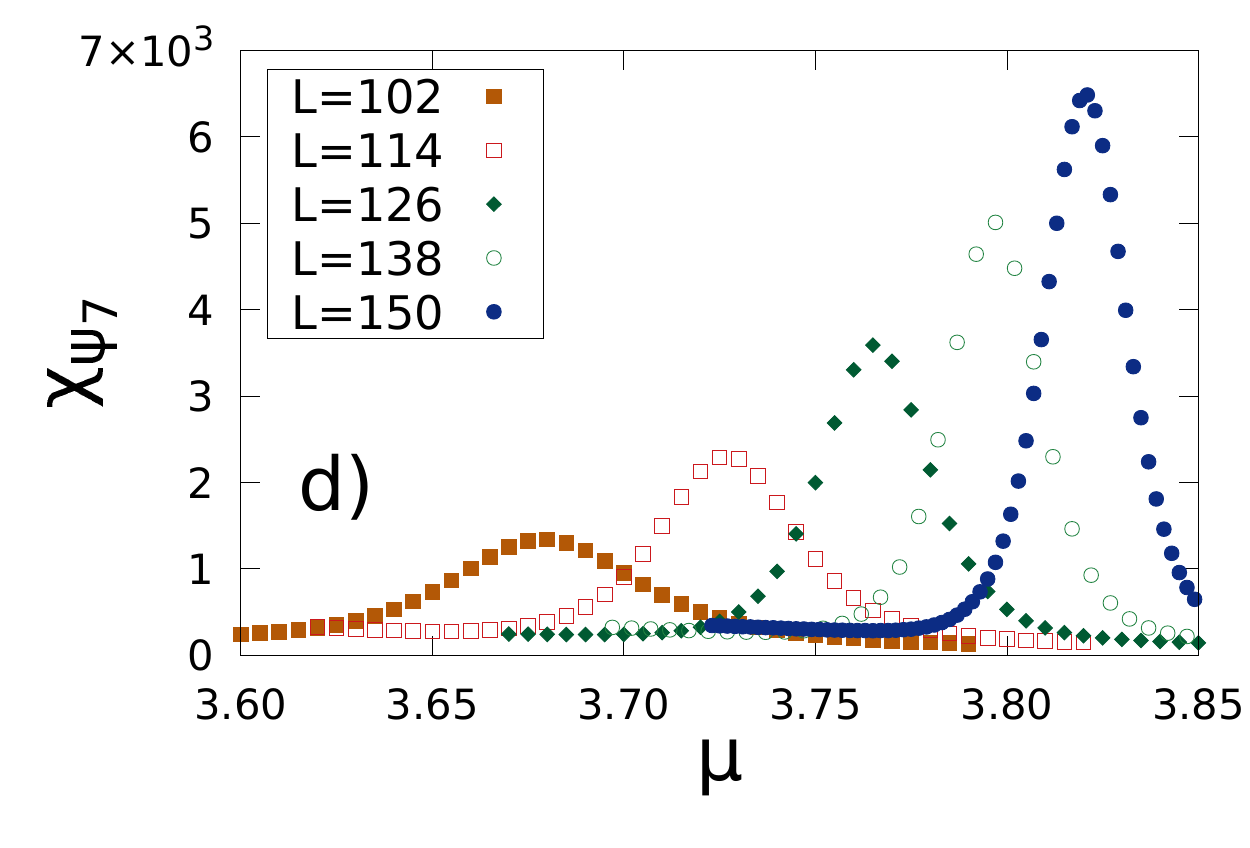}
    \includegraphics[width=.66\columnwidth]{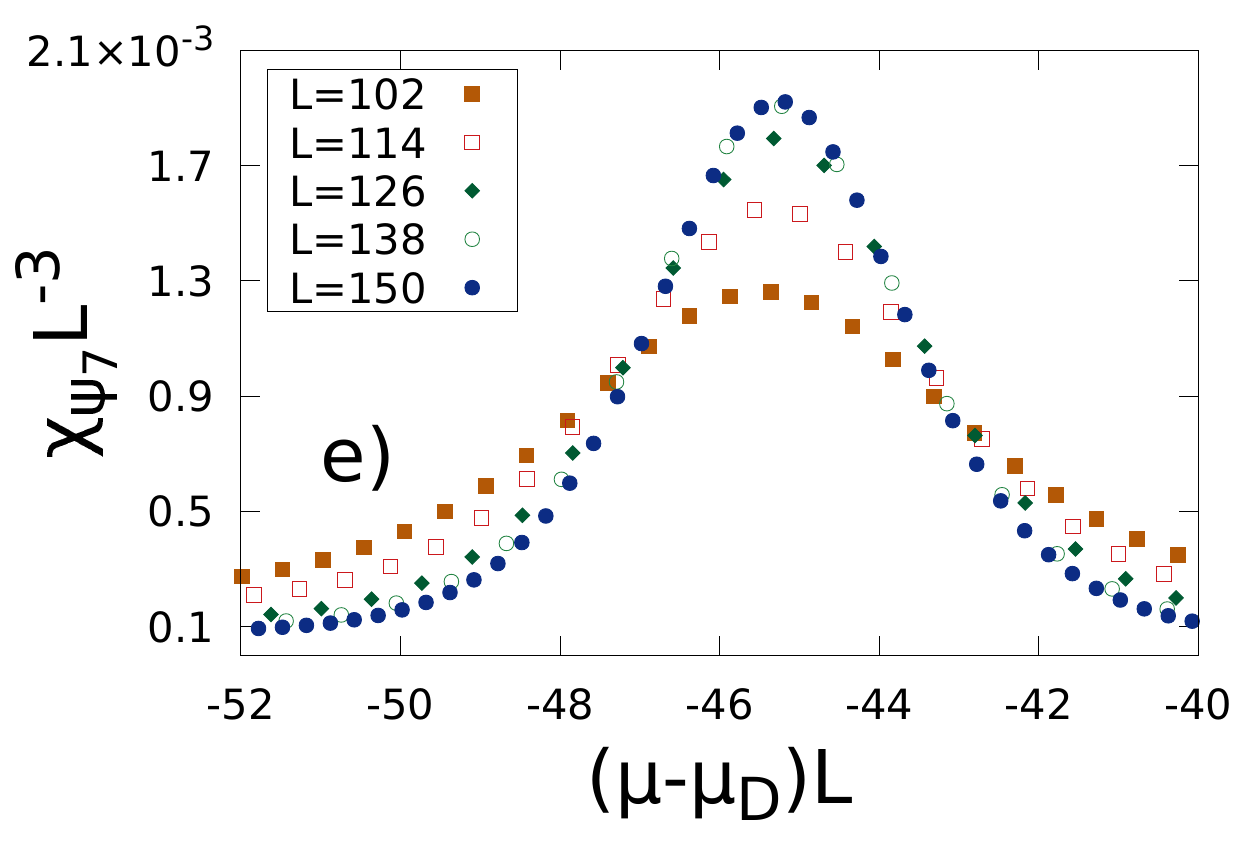}
    \includegraphics[width=.66\columnwidth]{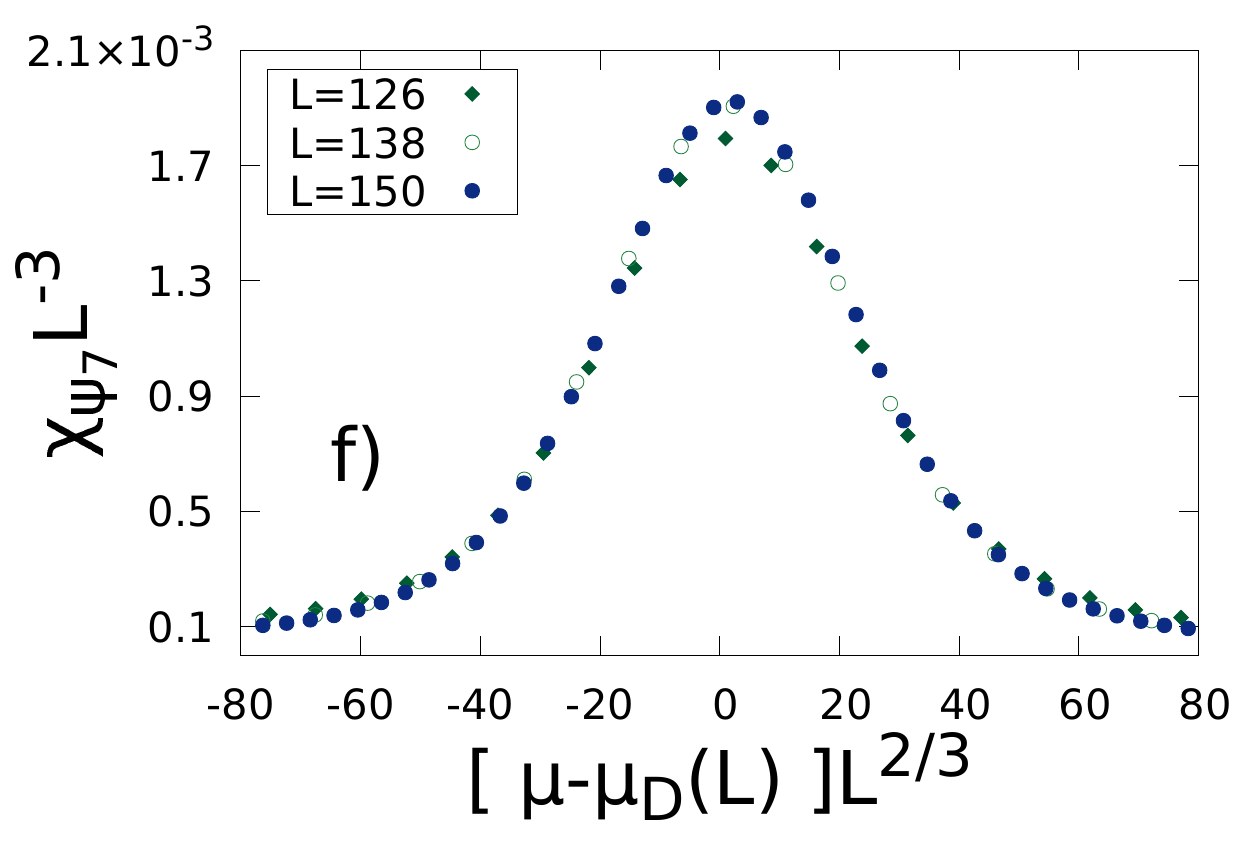}
    \caption{Results for (a) the order parameter $Q_{\psi_7}$ and (d) its susceptibility for the 5NN case as function of chemical potential. We find that $\mu_D(L)$ scales with $L^1$, but the width of susceptibility does not (b) and (e). We separate the scales by centering the curves around zero and re-scale with best-fitting exponents. We find in panels (c) and (f) that the width of susceptibility scales with $L^{2/3}$ and its maximum value with $L^3$. Since the histogram on the inset of panel (a) indicates a first order phase transition, these are non-standard scaling laws. The transition point used was $\mu_D=4.125$.}
    \label{qmu5nn_first}
  \end{center}
\end{figure*}

As can be seen in Fig.~\ref{qmu5nn}, panel (b), the order parameter $Q_5$ is not very sensitive in detecting this ordering within the system, specially for larger $L$. In order to characterize the phase behavior we notice that neighbors of order six are occupied in the bulk of domains and, whenever a neighbor of order seven is occupied, a domain boundary is formed. Therefore, we measure the occupancy of neighbors of order seven ($\psi_{7}$, similar to Eq.~\ref{eq:psi6}) and estimate the number of domain borders ($N_b$) of size $L$ containing $L/2$ particles $N_b = \frac{N\overbar{\psi_{7}}}{L}$, where $N$ is the total number of particles and $\overbar{\psi_{7}}$ is the lattice average value of $\psi_{7}$. Finally, we define the following order parameter

\begin{equation}
    Q_{\psi_7} = \frac{N_b}{6},
\end{equation}
which is greater than one if there are more than two domain boundaries of size $L$ per lattice direction and is equal to one if there are exactly two (Fig.~\ref{snap5nn}). Our simulations show two peaks in the histogram of this order parameter (inset of panel (a) in Fig.~\ref{qmu5nn_first}), indicating a first order phase transition. The results are shown in Fig.~\ref{qmu5nn_first}.

As can be seen, there are two separate scales in the finite size scaling analysis. First, in panels (b) and (d) we show that the location of the transition point $\mu_D(L)$ scales linearly with system size~($L^1$), with the phase transition in the thermodynamic limit occurring at $\mu_D=4.125$. Second, in same panels, we observe that the width of the curves of susceptibilities do not follow the same relation as the location of the transition point. In order to separate both scales, in panels (c) and (f) we center the curves of order parameter and susceptibility, respectively, around zero and re-scale with the best-fitting exponents. Both curves of order parameter and susceptibility scale with $L^{2/3}$ instead of $L^2$, as would be expected from a first order phase transition. Another non-standard result is that the maximum of susceptibility scales with $L^3$, as panels (e) and (f) show.

As chemical potential is increased the system undergoes a second phase transition, resulting in only one sublattice of Fig.~\ref{sl5nn} being occupied in the full packing configuration. Our simulations for this second transition show rare changes between ordered and disordered phases leading to poor sampling. We observed that using the Wang-Landau algorithm with adaptive windows~\cite{wlAdap} is more efficient than the one described in Sec.~\ref{sec:model}. The results shown in Fig.~\ref{qmu5nn} were obtained using this multicanonical sampling.

\begin{figure}[thb] 
  \begin{center}
    \includegraphics[width=.49\columnwidth]{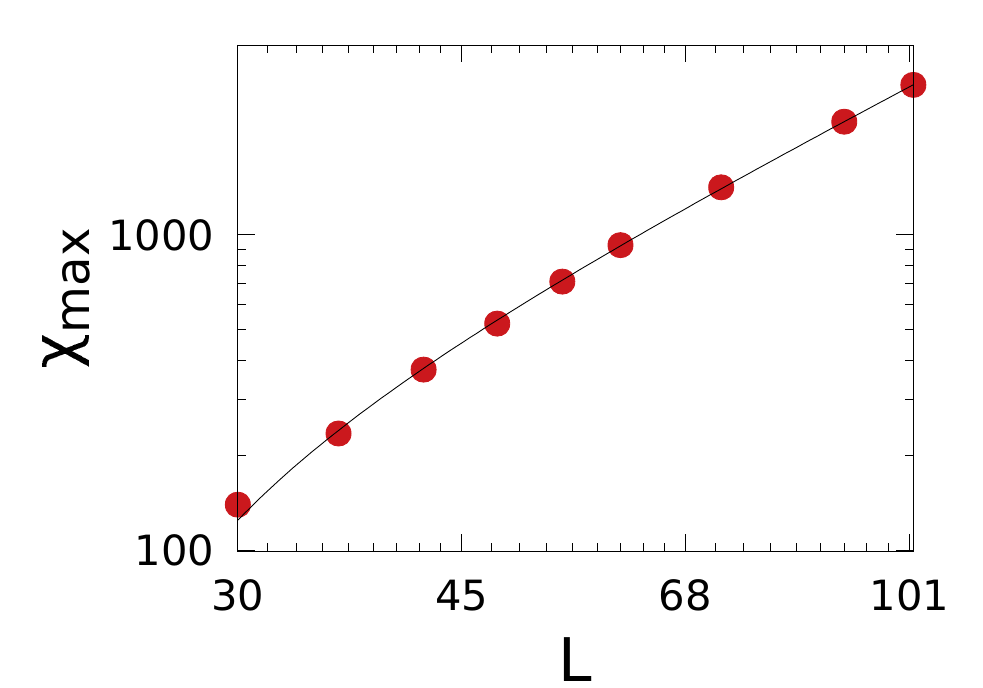}
    \includegraphics[width=.49\columnwidth]{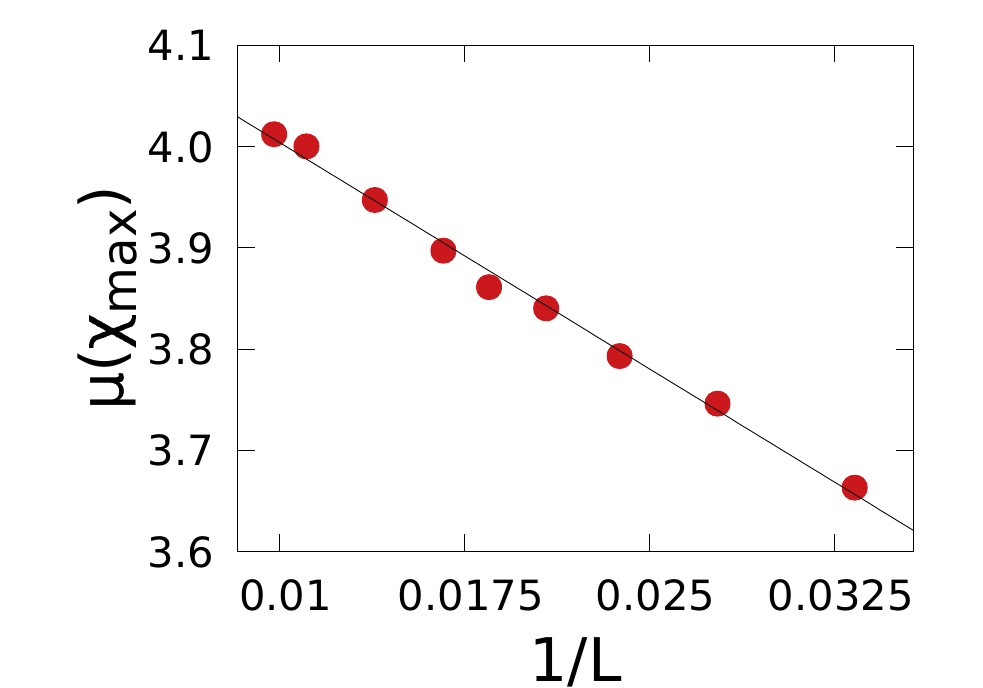}
    \caption{Finite size scaling analysis for the second phase transition ($Q_5$) in the 5NN case. On the left panel, by fitting $\chi_{\textrm{max}}$ to a power law we find the slope $2.13$ and, on the right panel, we see that $\mu (\chi_{\textrm{max}})$ scales with $L^1$ where a scaling with $L^2$ would be expected since it is a first order phase transition.}
    \label{fss5nn}
  \end{center}
\end{figure}

\begin{figure}[hbt] 
  \begin{center}
    \includegraphics[width=.8\columnwidth]{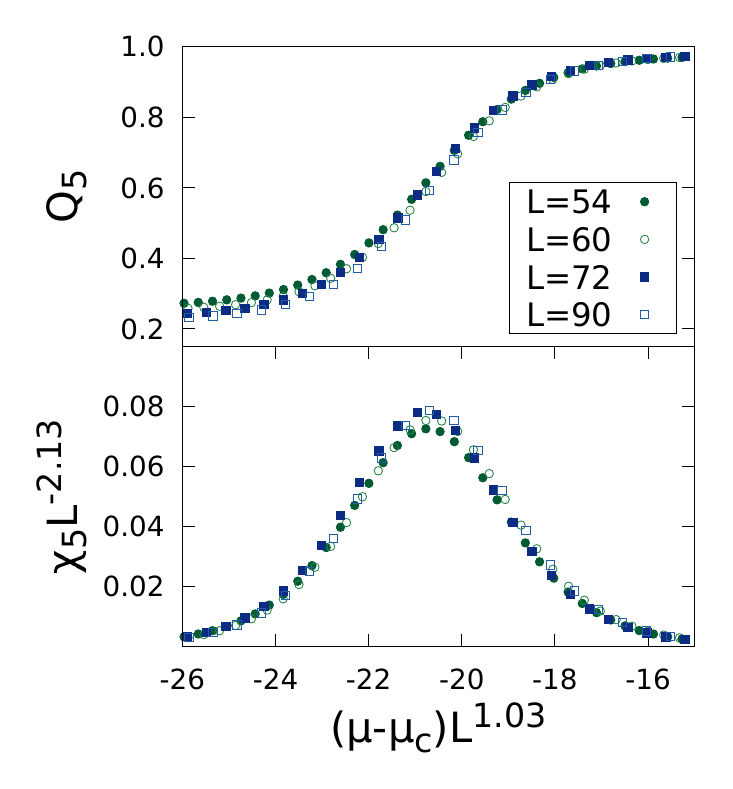}
    \caption{Finite size scaling collapse of the 5NN order parameter $Q_5$ (top) and its susceptibility (bottom) for different $L$ with scaling laws obtained in Fig.~\ref{fss5nn}. The transition point occurs at $\mu_{SL}=4.2$.
    }
    \label{col5nn}
  \end{center}
\end{figure}

Surprisingly, this transition to a sublattice phase also shows unusual scaling as can be seen through the finite size analysis in Figure~\ref{fss5nn}. We find that this phase transition occurs at $\mu_{SL}=4.2$ and present the collapse of curves in Fig.~\ref{col5nn}.

In a phenomenological study of the Ising model~\cite{binderDroplets}, it was found that, if stable domains are formed, the scaling laws for this first order phase transition are modified due to surface effects. Similar observations were observed in Monte Carlo simulations of the Baxter-Wu model~\cite{martinos2006}.
Since we observe stable domain formation in the 5NN model, the scaling laws we find could be related to this phenomena.

In summary, we found two phase transitions as chemical potential is increased. First, from a fluid-like phase, at $\mu_D=4.125$ the system organizes into two domains running along the three lattice directions. Second, a transition to a sublattice phase occurs at $\mu_{SL}=4.2$ and the full packing configuration is reached. Since the reasons for the non-standard scaling in this model are not completely clear, a theoretical approach or a different numerical method like the one described in~\cite{Fiore2013} would be of great value to improve the understanding of this behavior.

\section{\label{sec:conjec}$k$NN conjecture}

\begin{figure}[thb] 
  \begin{center}
    \includegraphics[width=.8\columnwidth]{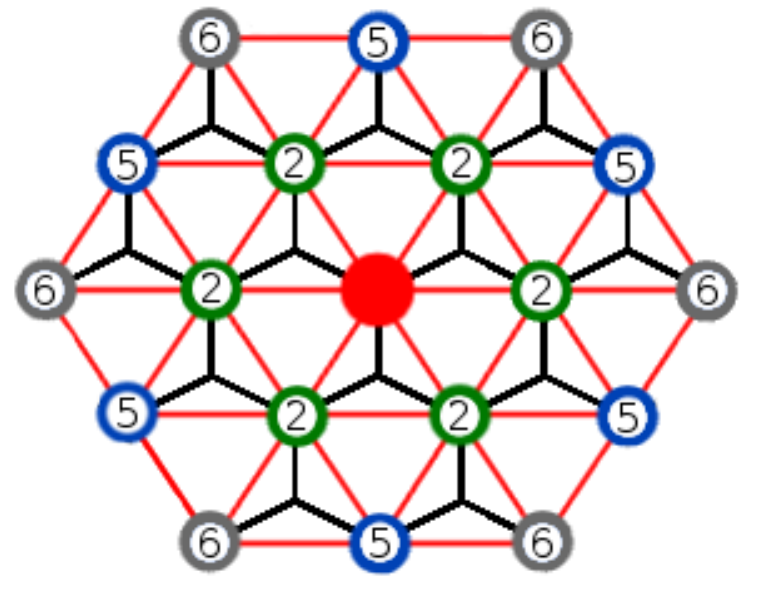}
    \caption{First three cases where condition (ii) is met. The full packing configuration of models $1$, $4$ and 5NN on the honeycomb lattice is the same as the cases $0$, $1$ and 2NN on the triangular lattice (condition (iii)).}
    \label{redeTriang}
  \end{center}
\end{figure}

\begin{figure}[thb] 
  \begin{center}
    \includegraphics[width=7.2cm]{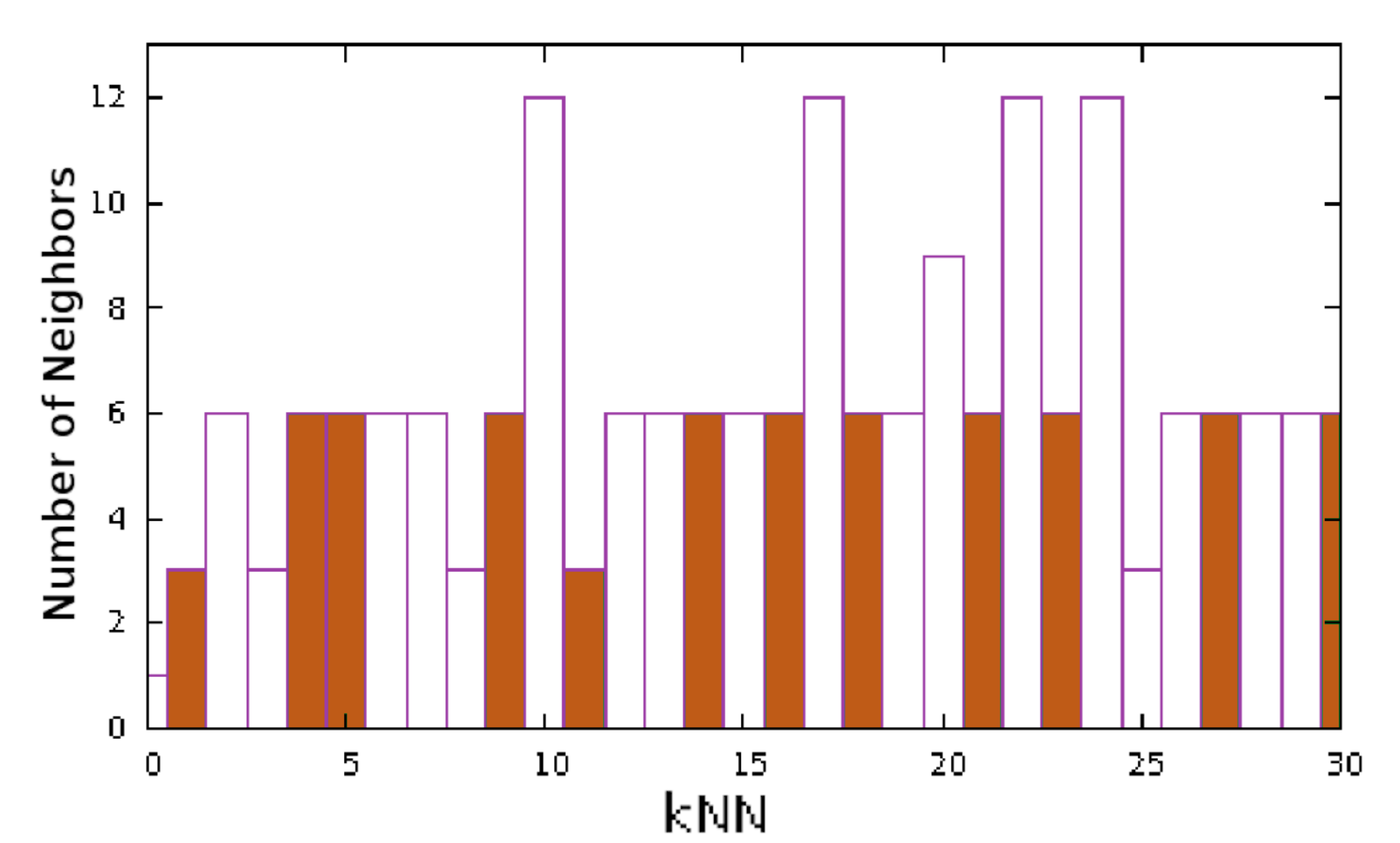}
    \caption{Number of neighbors of order $k$. Filled boxes show cases where condition (ii) is met and the full packing configuration on the honeycomb lattice allows only one type of site. The \textit{i-th} filled box relates to the $(i-1)$NN model on the triangular lattice. For clarity, we include only up to $k=30$, but computations for any $k$ is straightforward.}
    \label{histknn}
  \end{center}
\end{figure}

As we have seen in the models presented in this paper, some special values of $k$ are expected to show an $A$-$B$ phase transition with full packing similar to a related (but not equivalent) model on the triangular lattice. These cases are interesting because they may show more than one phase transition as density is increased. First, an ordered phase with both $A$ and $B$ particles may be formed, followed by an ordered phase with only one type of particle. In this section, we investigate which values of $k$ show this property in the following conjecture:

\begin{enumerate}
\item[i.] With an exclusion of up to $k$NN, if there is no sliding freedom, at full packing configurations neighbors of order $k+1$ should be preferentially occupied.
\item[ii.] If neighbors of order $k+1$ of a given $A(B)$ site are also $A(B)$ sites (Fig.~\ref{redeTriang}), the full packing configuration of model $k$NN contains only one type of particle and an $A$-$B$ phase transition is expected.
\item[iii.] Since each sublattice ($A$/$B$) forms a triangular lattice, the full packing configuration of the \textit{i-th} case satisfying condition (ii) on the honeycomb lattice (filled boxes in Fig.~\ref{histknn}) is related to the $(i-1)$NN model on the triangular lattice.
\end{enumerate}

Figure~\ref{histknn} shows the number of neighbors of order $k$ and cases where condition (ii) is met.

By performing quick simulations for cases $k=9$, $11$ and $14$, we checked that an $A$-$B$ transition is present in all of them.

Currently, we are expanding this conjecture using the full honeycomb lattice point group of symmetries in order to determine the high density phase of the $k$NN model on the honeycomb lattice. Using a different approach, the high density phases on the square lattice have been investigated~\cite{Nath_2016} and authors find a finite number of cases where a columnar phase is expected. On the honeycomb lattice, we expect some models to show either a columnar or hexatic phase (as the one observed in Ref.~\cite{darjani_2019} on the triangular lattice) transitioning into a sublattice phase while others transition from a fluid into a sublattice phase directly. Our results will be published in a future paper.

\section{\label{sec:summary}Summary and Conclusions}

In this paper we systematically studied high density phases and phase transitions in hardcore lattice gases on the honeycomb lattice. We performed Monte Carlo simulations of systems with exclusion region of up to fifth nearest neighbors, see Fig.~\ref{exclusion}, and proposed a conjecture concerning further exclusion areas. 
We observe a strong influence from underlying honeycomb lattice and symmetries of excluded regions in determining high density phases, with several interesting phenomena arising from the presence of asymmetrical particles and their high density packing. Non-standard scaling, columnar and high density domain-like phases are examples of the interesting aspects we observe in our investigations.
Due to large gaps in free energy between the phases observed, we employed several different techniques in order to efficiently sample the phase space, including cluster algorithms~\cite{rajesh} with sliding movements~\cite{yShapedRaj} and multicanonical Wang-Landau sampling with adaptive windows~\cite{wlAdap}. Even using these techniques, with our computational resources we were able to run simulations only on relatively small systems in cases 2NN ($L=108$) and 5NN ($L=102$). It is worth noting that these sizes should be compared to $L_{SQ}=\sqrt{2}L_{HC}\simeq 152$ and $L_{SQ}\simeq144$ on the square lattice since the honeycomb lattice has $N=2L^2$ sites. In other cases, where this strong slowing down is not observed, we performed simulations on systems with sizes up to $L=600$.

Our results show that, in the nearest neighbors exclusion case (1NN, Sec~\ref{subsec:1nn}), the system undergoes a second order phase transition at $\mu_c=2.064$, with critical exponents in the $2D$-Ising universality class. We present full finite size scaling analysis from data collapse, confirming previous predictions made by Runnels and Debierre~\cite{runnels1HC,debierre1HC} using matrix methods.

Systems with exclusion up to second nearest neighbors (2NN, Sec~\ref{subsec:2nn}) undergo a two step melting where a phase transition from a close-packed columnar phase to a solid-like domain phase is observed, followed by continuous shrinking of domains until fluid-like configurations are reached.
Although no inflection point in density is observed in passing from the fluid configurations into the solid-like phase, we characterize domain growth by defining a local order parameter where the occupation of all sixth nearest neighbors of a particle is tracked. We find that these neighbors are preferentially occupied in the domains phase, while almost none of them are occupied in the fluid configurations ($\mu\simeq 3.6$). We also observe that the maximum density reached in the solid-like phase strongly increases with system size, creating rigid configurations with slow dynamics that greatly reduce sampling efficiency.

As the system changes from the solid-like into the columnar phase, simulations show clear signs of a first order phase transition. From finite size scaling analysis, we find a non-standard relation predicted by a number of recent studies~\cite{JankeNonstandard,IsingPlaquettes,fssStuebel} where physical quantities scale with $L^1$ instead of the standard $L^2$. Whether this scaling is due to the close-packed configuration degeneracy exponentially growing with system size or due to surface interactions in the domains phase still not clear. We also observe a strong drift of the transitions region as the system size is increased, making it difficult to obtain the exact location of the transition in the thermodynamic limit.


Increasing the exclusion region up to third nearest neighbors (3NN, Sec~\ref{subsec:3nn}), we find no symmetry breaking at high densities ($\phi\simeq 0.98$). Short range local order is observed but no global order appears in our simulations. To further support the lack of phase transition we perform canonical simulations at a fixed density and use the argument of sublattice instability~\cite{yShapedRaj} to check that, indeed, a disordered phase is preferred over a columnar or solid-like phase at densities below $\rho_{max}$.

We also map this model into the problem of triangular trimers on the triangular lattice, which has an approximate analytical result at full packing configurations~\cite{verberkmoesTriang}. This model has a symmetry break in occupancy of up and down trimers as the density of domain boundaries is reduced. In order to reduce boundaries, two different chemical potentials are assigned to trimers, which is not the case in our model, where equal chemical potentials are assigned for $A$ and $B$ particles. Their model predicts no symmetry break in this regimen and, therefore, no phase transition should be observed as density is increased, which also corroborates our simulations. It should be stressed that, to our best knowledge, this is the first case where simulations of a hardcore model do not show any signs of phase transition even in packing fraction as high as $\phi=0.98$. Since this observation does not seem to be a finite size effect, a theoretical approach could shed some light on what is happening.

The 4NN model (Sec~\ref{subsec:4nn}) undergoes a second order phase transition at $\mu_c=2.6108$. We performed simulations on systems with sizes up to $L=420$ and provide numerical evidence on the nature of this phase transition. We observe that the $A$-$B$ symmetry break occurs slightly before ($\mu=2.607$) the sublattice transition but there are no significant differences in particles arrangement in the intermediary phase that could characterize these transitions as two separate critical points. By means of finite size scaling analysis, we obtain a set of critical exponents very close to the 3-state Potts model, in accordance with a continuous phase transition. This result corroborates the two observed peaks in histograms getting closer with increasing system size, where an increasing free energy gap would be expected in a discontinuous transition.

 When excluding neighbors of order up to $5$ (Sec.~\ref{subsec:5nn}), we find a strong first order phase transition at $\mu_c=4.2$. We use the Wang-Landau sampling with adaptive windows~\cite{wlAdap}, which has shown to be more efficient than the cluster algorithm used in previous cases. 
 This phase transition also shows non-standard scaling, where quantities scale with $L$ instead of $L^2$. We present the full scaling analysis and collapsed curves.
 
 As density is further decreased, we find a second discontinuous phase transition where the system organizes into domains of linear size $L$ running along all three lattice directions. We propose an order parameter which depends linearly on $L^{-1}$ and investigate the scaling behavior as the system size is increased. We find that the height of susceptibility scales with $L^3$, its width with $L^{2/3}$ and the location of critical point scales with $L$. We present the collapsed curves for the order parameter and its susceptibility.
 
 For further exclusion regions, we propose a conjecture concerning the possibility of more than one phase transition as density is decreased from a full packing configuration. This conjecture (Sec.~\ref{sec:conjec}) is based on geometrical arguments similar to those developed in Refs.~\cite{rajesh,yShapedRaj,ramola2012} and observation of systems extensively simulated in this paper with exclusion regions up to 5NN. Quick simulations on cases with $k=9$, $11$ and $14$ confirmed our conjecture predictions of a symmetry break in the occupancy $A$ and $B$ sites at high densities as well as the prediction of a hexatic phase on the case with exclusion up to ninth nearest neighbors ($k=9$). We are currently expanding this conjecture and will publish our results in a future paper.
 
A more formal approach, as series expansions from ordered phases developed in Refs.~\cite{rajesh,yShapedRaj,ramola2012}, could help us better understand the exact origin of phase instability and their effects in thermodynamic properties. In Ref.~\cite{yShapedRaj}, the authors comment on the difficulties of performing series expansion in systems on the honeycomb
lattice, some of them originated by the presence of two types of sites on this lattice, which should be treated separately. 

As a final remark, we point to the question on the general applicability and efficiency of the cluster algorithm developed in Ref.~\cite{rajesh}.
Even though this algorithm, with the aid of sliding movements, helped us to improve sampling during simulations, we observed a relative poor performance in systems where a much slower dynamics (2NN and 5NN) is observed, at least when compared to other systems studied in this work. In Ref.~\cite{ramola2015}, authors apply a generalization of the cluster movement to a mixture of plaquettes and squares and note that the generated dynamics do not remove winding sectors at domain boundaries. Maybe, this could be the case here.

\section{Acknowledgments}

F.C.T. thanks the Brazilian Agency CNPq for its financial support. H.C.M.F acknowledges the Universitat de Barcelona during his stay.



\bibliographystyle{apsrev4-1}

\end{document}